\documentclass[twocolumn,journal]{IEEEtran}
\usepackage{verbatim}
\usepackage{amsmath}
\usepackage{amsthm}
\usepackage{amssymb}
\usepackage{color}
\usepackage{bm}
\usepackage{multirow}

\usepackage[pdftex]{graphicx}

\newtheorem{proposition}{Proposition} 
\newtheorem{theorem}{Theorem}
\newtheorem{corollary}{Corollary}
\newtheorem{remark}{Remark}
\newtheorem{lemma}{Lemma}

\newtheorem{example}{Example}

 \newenvironment{proofof}[1]{\vspace*{5mm} \par \noindent
         \quad{\it Proof of #1: \hspace{2mm}}}{\endproof
}

\def\FF{\mathbb{F}}

\def\argmax{\mathop{\rm argmax}}

\def\mincut{\mathop{\rm mincut}\nolimits}

\def\Label#1{\label{#1}\ [\ \text{#1}\ ]\ }
\def\Label{\label}

\begin{document}

\title{Secure Network Code for Adaptive and Active Attacks with No-Randomness in Intermediate Nodes
\thanks{
MH 
was supported in part by 
the JSPS (Japan Society for the Promotion of Science) Grant-in-Aid for Scientific Research 
(A) No.17H01280, (B) No. 16KT0017, (C) No. 16K00014, 
and Kayamori Foundation of Informational Science Advancement.}}

\author{
Ning Cai, \IEEEmembership{Fellow, IEEE} and
Masahito Hayashi, \IEEEmembership{Fellow, IEEE}
\thanks{
Ning Cai is with the School of Information Science and Technology, ShanghaiTech University
(e-mail: ningcai@shanghaitech.edu.cn).
Masahito Hayashi is with the Graduate School of Mathematics, Nagoya University, Furocho, Chikusa-ku, Nagoya, 464-8602, Japan. 
He is also with 
Shenzhen Institute for Quantum Science and Engineering, Southern University of Science and Technology,
No.1088 Xueyuan Avenue, Nanshan District, Shenzhen, 518055, China,
the Centre for Quantum Technologies, National University of Singapore, 3 Science Drive 2, 117542, Singapore,
and
Center for Quantum Computing, Peng Cheng Laboratory, Shenzhen 518000, China
(e-mail:masahito@math.nagoya-u.ac.jp).
}}

\markboth{N. Cai and M. Hayashi: Secure Network Code for Adaptive and Active Attacks}{}

\maketitle

\begin{abstract}
In secure network coding, there is a possibility that
the eavesdropper can improve her performance
when 
she changes (contaminates) the information on the attacked edges (active attack)
and chooses the attacked edges adaptively (adaptive attack).
We analyze the security for network code over such types of attacks.
We show that active and adaptive attacks cannot improve the performance of the eavesdropper when the code is linear.
Further, we give a non-linear example, in which 
an adaptive attack improves the performance of the eavesdropper.
We derive the capacity for the unicast case and the capacity region for the multicast case or the multiple multicast case in several examples of relay networks, beyond the minimum cut theorem, 
when no additional random number is allowed as scramble variables in the intermediate nodes.
No prior study compared 
the difference of the capacity and the capacity region 
between the existence and the non-existence of randomness in the intermediate nodes 
under these network models
even with non-adaptive and non-active attacks.
\end{abstract}


\section{Introduction}\Label{S1}
Secure network coding is a method securely transmitting
information from the authorized sender to the authorized receiver. 
Cai and Yeung \cite{Cai2002,CY,YN} discussed the secrecy for the malicious
adversary, Eve, wiretapping a subset $E_E$ of all channels in the network. 
The papers \cite{CG,RSS,FMSS,NYZ,HY,Cai,CC,AVF16-1,AVF16-2,AVF17} developed several types of secure network coding.
In particular, the papers \cite{CG,AVF16-1} considered security for multiple unicast sessions within the context of network coding.

Combining the codes in \cite{Cai2002} and \cite{CY2},
the paper \cite{NY} proposed a linear code to protect transmission from attacks of eavesdrop and injection of error (contamination) simultaneously.
Like traditional error correcting code and error correction network code (i.e. against Byzantine attack) in \cite{YC2,CY2}, 
the paper \cite{NY} considered 
the robustness in the worst case, or equivalently 
it evaluated the error probability when the adversary to inject error knows the message to be sent. 
However, it discussed the secrecy independently of the robustness, i.e.,
it considered the secrecy only when the information on the network is not changed.
Also, the papers \cite{SK,KMU,Matsumoto2011,Matsumoto2011a} showed the existence of a secrecy code that
universally works for any types of eavesdroppers under the size
constraint of $E_E$. 
In particular, the papers \cite{Matsumoto2011,Matsumoto2011a} constructed it by using the universal hashing lemma \cite{bennett95privacy,HILL,hayashi11}.
Further, the papers \cite{SK,KMU,Yao2014} evaluated errors 
only when the information on a part of network is changed,
but they evaluated the secrecy only when the information on the network is not changed
or Eve did not know the replaced information.

On the other hand, as another model, we assume that the goal of Eve to inject error is to help having more information about the message. 
In this case, she may inject error according to the knowledge which she obtained from her previous action but not the message. 
This improvement for her ability of eavesdropping is an essential difference between the two models.
The recent paper \cite{HOKC} discussed this model, i.e.,
evaluated the secrecy as well as the error 
when Eve contaminates the eavesdropped information and knows the replaced information. 
(For the detailed relation, see Remark \ref{R9}.)

The effects of Eve's contamination depend on the type of the network code.
It is well-known that linear network code is optimal for single source network \cite{LYC}. 
But in many cases, linear codes are not optimal, or in other words, non-linear code has better performance, for example, in coding for multiple source network and classical error correcting code (which can be considered as error correction network coding for a two-node network, 
``point-to-point network)''. 
In several known examples of multiple source network, 
non-linear code can do better than linear one 
\cite{DFZ}.
As matter of fact, linear code has many advantages, e.g., easy for handle, lower complexity of encoding and decoding etc. 
When the code is linear, as shown in \cite{HOKC},
any contamination (any active attack) does not improve her performance.
However, when the code is not linear, there exists only one example where the 
contamination improves her performance \cite{HOKC}.

Despite these developments, there are still some problems in existing studies.
Although these existing studies achieved the optimal rate with secrecy condition,
their optimality relies on the minimum cut theorem.
That is, they assumed that the eavesdropper may choose any $r$-subset channels to access, and did not address another type of conditions for the eavesdropper.
For example, the studies \cite{SK,KMU,Matsumoto2011,Matsumoto2011a} optimized only the codes in the source and terminal nodes
and did not optimize the coding operations on the intermediate nodes.
Also, in other existing studies, 
the intermediate nodes do not have as complicated codes as the source and terminal nodes.
In this paper, to achieve the optimal rate beyond the minimum cut theorem,
we address the optimization of the coding operations on the intermediate nodes as well as on the source and terminal nodes.

Further, we consider a new type of attacks, adaptive attacks.
Assume that distinct numbers are assigned to the edges, 
and the communication on edges are done in the decreasing order for 
the assigned numbers.
Usually, Eve cannot decide the edges to be attacked depending on the previous observation.
Now, we allow Eve to choose the edges to be attacked 
based on the previous observations. 
Indeed, the channel discrimination,
it is known that such an adaptive strategy does not improve the asymptotic performance \cite{Channel}.
Then, we find two characteristics for adaptive attacks, which are similar to the case of active attacks.
First, we find a non-linear code where an adaptive attack significantly improves Eve's performance.
Using this characteristic, we find an example of a non-asymptotic network model, which has 
no secure code for adaptive attacks, but has 
secure code for conventional attacks.
Second, we show that any adaptive attack cannot improve Eve's performance when the code is linear.
Using this fact, we derive the asymptotic performance in several typical network models in the following way
when Eve is allowed to use adaptive and active attacks.

In this paper, we discuss the asymptotic securely transmittable rate over the above attacks
not only for a unicast network but also for
a multiple multicast network, in which,
multiple senders are intended to send their different messages to different multiple receivers.
Under these settings, we define 
the capacity and the capacity regions for given network models,
and calculate them in several examples.
For the definition, we define two types of capacity regions depending on 
the requirement on the code on the intermediate nodes.
Usually, a secure network code employs scramble random numbers, 
which need to be physical random numbers different from pseudo random numbers.
In the first capacity region, we allow each node to introduce new scramble random numbers unlimitedly.
Here, the scramble random numbers of each node are not shared with other nodes 
and should be independent of random variables in other players and other nodes
before starting the transmission.
In the second capacity region, only source nodes are allowed to employ scramble random numbers
due to the following reason.
To realize physical random numbers as scramble random numbers, we need a physical device.
If the physical random number has sufficient quality, the physical device is expensive and/or 
consumes a non-negligible space because
it often needs high level quantum information technologies with advanced security analysis \cite{HG,HZ}.
It is not so difficult to prepare such devices in the source side.
However, it increases the cost to prepare devices in the intermediate nodes
because networks with such devices require more complicated maintenance than a conventional network.
Therefore, from the economical reason, it is natural to impose this constraint to our network code.
Unfortunately, only a few papers \cite{CY-07,CHK-10,CHK-13} discussed such a restriction.
Hence, this paper addresses the difference between the capacities with and without such a restriction by introducing 
the no-randomness capacity and the full-randomness capacity.
Further, as an intermediate case, 
by introducing the limited-randomness capacity,
we can consider the case when the number of available scramble random numbers in each intermediate node
is limited to a certain amount.
Then, the relation between our capacities and the existing studies is summarized as Table \ref{TX}.
In addition, for both types of capacities and capacity regions, 
we define the linear codes version, in which, our codes are limited to linear codes.
We also show that the linear version of capacities and capacity regions
are the same as the original capacities and capacity regions under the above examples
because the optimal rate and rate regions in the original setting can be attained by linear codes.

The remaining parts of this paper is organized as follows.
Section \ref{S4} gives the formulation of our network model. 
Section \ref{S5} gives an example of network model, in which, an adaptive attack efficiently improves Eve's performance.
To discuss the asymptotic setting,
Section \ref{S5B} defines the capacity region.
Section \ref{S6} discusses the relay network model and derives its capacity.
Section \ref{S8} discusses the homogenous multicast network model and derives its capacity.
Section \ref{S9} discusses the homogenous multiple multicast network model and derives its capacity.
In Section \ref{S7}, we give an important lemma, which is used in the converse part in the above models.
\begin{table*}[!t]
  \caption{Summary of comparison with existing results}
\label{TX}
\begin{center}
  \begin{tabular}{|l|c|c|c|c|} 
\hline
& Active & Adaptive &  Node   & \multirow{2}{*}{Linearity} \\
& attack & attack  &  randomness&\\
\hline
Papers \cite{Cai2002,CY,YN,FMSS} 
& not allowed & not allowed &  not allowed & scalar \\
Paper \cite{CG}
& not allowed & not allowed &  allowed & non-linear \\
Papers \cite{Cai,CC,CY-07,CHK-10,CHK-13}
& not allowed & not allowed &  allowed & scalar \\
Papers \cite{RSS,HY} & not allowed & not allowed &not allowed  & scalar \\
Papers \cite{Matsumoto2011,Matsumoto2011a,SK} & not allowed & not allowed &not allowed  & vector \\
Papers \cite{KMU,Zhang} & semi active attack & not allowed &not allowed  & vector \\
Paper \cite{AVF16-1} & not allowed & not allowed & not allowed& (*1) \\
Papers \cite{AVF16-2,AVF17} &not allowed & not allowed& not allowed& scalar \\
\multirow{2}{*}{Paper \cite{HOKC}}& 
\multirow{2}{*}{allowed} & \multirow{2}{*}{not allowed}  & 
\multirow{2}{*}{not allowed} & vector/ \\
&&&& non-linear\\
\hline
Our non-linear example & not allowed & allowed & not allowed & non-linear \\
No-randomness capacity & allowed & allowed &  not allowed& vector \\
Limited-randomness capacity & allowed & allowed  & partially 
& vector \\
Full-randomness capacity & allowed & allowed & allowed 
& vector \\
\hline
  \end{tabular}
\\
 \end{center}
\vspace{2ex}
Node randomness expresses the random number generated in intermediate nodes,
which is independent of the variables in other nodes and other players before starting the 
transmission.
Linearity expresses whether the code is linear or not.
When it is linear, the column
expresses which linearity condition is imposed, 
scalar or vector linearity.
These two kinds of linearity conditions are explained in Section \ref{R2}.
Semi active attack means that 
Eve injects the noise in several nodes and eavesdrops several nodes,
but she estimates the message only from the eavesdropped information on the node
without use of the information of the noise. 
For the detailed relation for active attack,
see Remark 8 of \cite{HOKC}.
(*1) Paper \cite{AVF16-2} considers the channel that destroys linearity.
Hence, it does not care linearity. 
\end{table*}

\section{Adaptive and active attack for general network}\Label{S4}
\subsection{Formulation and reduction to non-adaptive attack}\Label{S4-1}
Now, we give the most general formulation of network coding and adaptive and active attacks.
We consider an acyclic general network with multiple multicast setting as follows.
The network has $a$ source nodes, $b$ terminal nodes,
several intermediate nodes, and $\ell$ edges,
where each edge is assigned to a distinct number  from $[\ell]:= \{1, \ldots, \ell\}$.
Hence, $[\ell]$ can be regarded as the set of edges.
Each edge transmits a single letter on a finite set ${\cal X}$.
Our task is the following.
The $i$-th source node securely sends the message $M_{i,j}$ to the $j$-th terminal node, where the messages are subject to independent uniform distribution.
Here, the tuple of all messages are denoted by $\bm{M}$.

Next, we assume that as scramble random numbers, 
each intermediate node can use additional uniform random numbers, which are independent of other random variables. They might be realized as physical random numbers.
The $i$-th source node converts the pair of the messages $(M_{i,1}, \ldots, M_{i,b})$
and the scramble random numbers to the tuple of the letters on the outgoing edges.
Each intermediate node converts the pair of the letters on the incoming edges and the scramble random numbers to the tuple of the letters on the outgoing edges.
The $j$-th terminal node converts the pair of the letters on the incoming edges to the tuple of
the recovered messages $(\hat{M}_{1,j}, \ldots, \hat{M}_{a,j})$.
We denote a network code by $\Phi$.
We denote the cardinality of the message $M_{i,j}$ by $|\Phi|_{i,j}$.
When $a=1$, we simply denote it by $|\Phi|_{j}$.
In particular, when $a=b=1$, we simply denote it by $|\Phi|$.
We denote the set of codes by $\mathfrak{C}^0$.

Now, we consider two conditions for our network code $\Phi$.
\begin{description}
\item[(C1)] [Linearity]
Any message, any scramble random number, and
information on any edge can be given as elements of vector spaces over the finite field $\FF_q$.
All of the conversions in source, intermediate, and terminal nodes are linear over $\FF_q$, i.e., they are written as matrices whose entries are elements of $\FF_q$.
Then, the code is called {\it linear} with respect to $\FF_q$\footnote{This type of linear code is often called vector linear \cite{Dougherty} because 
these random variables are given as elements of vector spaces over the finite field $\FF_q$.
Although the paper \cite{Dougherty} assumes that
all the messages, the scramble random numbers, and the variables on the edges have the same dimension,
we do not assume this condition.}.

Here, to apply the linearity condition, 
we choose a subset of ${\cal X}$ whose cardinality is a power of $q$.
Then, the information on any edge can be given as an element of vector space over the finite field $\FF_q$.
While all edges sent the information on the same set ${\cal X}$,
the above subset might depend on the edge.
This is because the dimension of the information to be sent depends on the edge in general.
Since the cardinality of the set ${\cal X}$ is an arbitrary number, 
we can apply this linearity condition to the case when ${\cal X}$ is a given as the $n$-th power of a certain set.

\item[(C2)] [No-randomness]
All of intermediate nodes have no scramble random numbers.

\item[(C2')] [Limited-randomness]
Each limited intermediate node has limited scramble random numbers.
When each group is composed of one node,
as a typical example, we assume that the node in $i$-th group can use $\gamma_i$ random numbers per transmission.
\end{description}

Next, we define Eve's attack.
The conventional attack is modeled by a collection $\mathfrak{A}^0$ of subset of $[\ell] $.
That is, in the conventional attack, 
Eve chooses a subset $\bm{s}\in \mathfrak{A}^0$,
and eavesdrops the edges in the subset $\bm{s}$.
This types of attack is called a {\it deterministic attack}.
Hence, the set of deterministic attacks is identified with $\mathfrak{A}^0$.
The following discussion depends on the 
collection $\mathfrak{A}^0$ of subset of $[\ell] $.
That is, our problem is characterized by the structure of network and 
the collection $\mathfrak{A}^0$. 
Also, Eve can randomly choose her choice $\bm{s}$.
Such an attack is written as a probability distribution $P_{\bm{S}}$
and is called a {\it randomized attack} or {\it a randomization of} $\mathfrak{A}^0$.
We denote the set of randomized attacks by 
$\bar{\mathfrak{A}}^0$.

In this paper, we allow Eve to adaptively choose 
the edges to be eavesdropped.
For simplicity, we assume that all subsets in the collection 
${\mathfrak{A}}^0$ have the same cardinality $\zeta$.
While Eve is allowed to eavesdrop $\zeta$ edges,
she can adaptively choose them as follows.
She chooses the first edge $\alpha_1 \in [\ell]$ to be eavesdropped,
and obtains the information $Z_1 \in {\cal X}$ on the edge.
Based on the information $Z_1$, she chooses the second 
edge $\alpha_2(Z_1) \in [\ell]$ to be eavesdropped
and obtains the information $Z_2 \in {\cal X}$ on the edge.
In this way, 
based on the information $Z_1, \ldots, Z_{j-1}$, she chooses the $j$-th
edge $\alpha_j(Z_1,\ldots, Z_{j-1}) \in [\ell]$ to be eavesdropped
and obtains the information $Z_j \in {\cal X}$ on the edge.
Since the choice of the set
$\bm{\alpha}=\{\alpha_1, \ldots, \alpha_{\zeta}\}$ of attacked edges
is given as a function of $ \zeta-1$ outcomes $z_1, \ldots, z_{\zeta-1}$,
it is often written as 
$\bm{\alpha}(z_1, \ldots, z_{\zeta-1})$ to clarify this point.
Here, for any data $z_1, \ldots, z_{\zeta-1}$, 
$\bm{\alpha}(z_1, \ldots, z_{\zeta-1})$
is required to belong to the family $\mathfrak{A}^0 $.
This type of attack is called a {\it general adaptive attack}.
In this type of attack, the order of eavesdropped edges has no relation with the numbers assigned to the edges.
A general adaptive attack $\bm{\alpha}=(\alpha_1, \ldots, \alpha_{\zeta})$
is called a {\it time-ordered adaptive attack}
when $\alpha_1< \alpha_2(z_1)< \ldots < \alpha_{\zeta}(z_1, \ldots, z_{\zeta-1})$.
Although a general adaptive attack has less practical meaning than
a time-ordered adaptive attack, 
we consider a general adaptive attack due to its mathematical simplicity.
We denote the sets of 
time-ordered adaptive attacks and general adaptive attacks by 
$\mathfrak{A}^1$ and $\mathfrak{A}^2$, respectively.
Similar to $\bar{\mathfrak{A}}^0$,
the sets of their randomizations are written as 
$\bar{\mathfrak{A}}^1$ and $\bar{\mathfrak{A}}^2$, respectively.
Now, we identify the set of deterministic attacks with the collection $\mathfrak{A}^0$.
Considering a constant function $\bm{\alpha}$, which does not depend on 
$ \zeta-1$ outcomes $z_1, \ldots, z_{\zeta-1}$,
we can consider the collection $\mathfrak{A}^0$ as a subset of $\mathfrak{A}^1$
while $\mathfrak{A}^1 \subset \mathfrak{A}^2$.

Next, we consider a more powerful attack than 
a time-ordered adaptive attack 
$\bm{\alpha}=(\alpha_1, \ldots, \alpha_{\zeta})$.
Although Eve decides the eavesdropped edges in the same way as
the time-ordered adaptive attack 
$\bm{\alpha}=(\alpha_1, \ldots, \alpha_{\zeta})$,
she is allowed to change the information $Z_j$ on the $j$-th eavesdropped edge $\alpha_j(Z_1, \ldots, Z_{j-1})$
to $\beta_{j}(Z_1, \ldots, Z_{j})$, 
which is a function of her observations $Z_1, \ldots, Z_{j}$.
This kind of attack is called an {\it adaptive and active attack}
and is written as
the pair $(\bm{\alpha},\bm{\beta})$ of 
$\bm{\alpha}=(\alpha_1, \ldots, \alpha_{\zeta})$
and 
$\bm{\beta}=(\beta_1, \ldots, \beta_{\zeta})$.
We denote the set of adaptive and active attacks (such functions) 
by $\mathfrak{A}^3$.
The sets of the randomizations 
are written as 
$\bar{\mathfrak{A}}^3$.
When $\bm{\alpha}$ does not depends on her observations $Z_1, \ldots, Z_{\zeta-1}$,
$\bm{\alpha}$ is a deterministic attack and 
the pair $(\bm{\alpha},\bm{\beta})$ is called an {\it active attack}.
Indeed, when active attack is made, the information on the network is changed.
However, in this paper, we do not care about the correctness of the recovered information
when active attack is made.
We consider the correctness in the decoding only when no active attack is made, i.e.,
we discuss only the secrecy when active attack is made.

Hence, we have the relations
$\mathfrak{A}^0\subset \mathfrak{A}^1\subset \mathfrak{A}^2,
\mathfrak{A}^0 \subset \mathfrak{A}^1\subset \mathfrak{A}^3$,
and 
$\bar{\mathfrak{A}}^0\subset \bar{\mathfrak{A}}^1\subset \bar{\mathfrak{A}}^2,
\bar{\mathfrak{A}}^0 \subset \bar{\mathfrak{A}}^1\subset 
\bar{\mathfrak{A}}^3$.
We also assume that there is no error in any edges except for the eavesdropped edge.
The classes of attacks are summarized as Table \ref{TX2}.

\begin{table}[htpb]
  \caption{Summary of classes of attacks}
\label{TX2}
\begin{center}
  \begin{tabular}{|l|c|} 
\hline
& Type of attacks \\
\hline
$\mathfrak{A}^0$ 
& deterministic attacks\\
$ \mathfrak{A}^1$ 
& time-ordered adaptive attack \\
$ \mathfrak{A}^2$ 
&  general adaptive attacks \\
 $ \mathfrak{A}^3$
 & adaptive and active attack \\
 $\bar{\mathfrak{A}}^0$
 & Randomizations of  ${\mathfrak{A}}^0$\\
 $\bar{\mathfrak{A}}^1$
 & Randomizations of  ${\mathfrak{A}}^1$\\
 $\bar{\mathfrak{A}}^2$
 & Randomizations of  ${\mathfrak{A}}^2$\\
 $\bar{\mathfrak{A}}^3$
 & Randomizations of  ${\mathfrak{A}}^3$\\
 \hline
  \end{tabular}
\\
 \end{center}
\vspace{2ex}
\end{table}

Under a code $\Phi$ and an attack $(\bm{\alpha},\bm{\beta}) \in \mathfrak{A}^3$,
we denote the mutual information between the messages and Eve's observations 
$\bm{Z}=(Z_1, \ldots, Z_{\zeta})$
by $I(\bm{M};\bm{Z})_{\Phi,(\bm{\alpha},\bm{\beta})}$.
Also,
under an attack $\bm{\alpha} \in \mathfrak{A}^2$
we denote it 
by 
$I(\bm{M};\bm{Z})_{\Phi,\bm{\alpha}}$.
In addition, 
an attack $P \in \bar{\mathfrak{A}}^i$ with $i=0,1,2,3$,
we denote it by $I(\bm{M};\bm{Z})_{\Phi,P}$.
Then, for any attack $P \in \bar{\mathfrak{A}}^i$ for $i=0,1,2,3$
and a network code $\Phi$, we can choose an attack 
$\bm{x} \in {\mathfrak{A}}^i$
such that $I(\bm{M};\bm{Z})_{\Phi,P}\ge  
I(\bm{M};\bm{Z})_{\Phi,\bm{x}}$.
That is, we have
\begin{align}
\max_{P \in \bar{\mathfrak{A}}^i}
I(\bm{M};\bm{Z})_{\Phi,P}
=
\max_{\bm{x} \in \mathfrak{A}^i}
I(\bm{M};\bm{Z})_{\Phi,\bm{x}}
\Label{ER}
\end{align}
 for $i=0,1,2,3$.

First, we consider the case when the network code is not necessarily linear. 
Then, we have the following theorem\footnote{
Even when the cardinality $d$ of each channel is different from $q$,
this theorem still holds.}
when $Y_i$ expresses the information on the edge $i$.

\begin{theorem}\Label{T7B}
Assume that a network code $\Phi$ satisfies the following condition.
Given an arbitrary element 
$\bm{s}=\{s_1,\ldots,s_\zeta\} \in 
 \mathfrak{A}^0$, we have
\begin{align}
H(\bm{M}| Y_{s_1}=z_{1}, \ldots, Y_{s_\zeta}=z_{\zeta})
=
H(\bm{M}| Y_{s_1}, \ldots, Y_{s_\zeta}) 
\Label{H8-17B}
\end{align}
for any element $(z_{1}, \ldots, z_{\zeta}) $.
Then, any general adaptive attack $\bm{\alpha} \in \mathfrak{A}^2$ satisfies 
\begin{align}
I(\bm{M};\bm{Z})_{\Phi,\bm{\alpha}} \le
\max_{ \bm{s} \in \mathfrak{A}^0}
I(\bm{M}; \bm{Z})_{\bm{s}}.\Label{H8-17}
\end{align}
\hfill $\square$\end{theorem}
Theorem \ref{T7B} will be shown in the next subsection.
Since 
$I(\bm{M}; \bm{Z})_{\Phi,\bm{s}}=0$ for any $\bm{s} \in  \mathfrak{A}^0$
implies the condition \eqref{H8-17B},
we have the following corollary.

\begin{corollary}\Label{T7C}
When the relation 
\begin{align}
I(\bm{M}; \bm{Z})_{\Phi,\bm{s}}=0 \Label{H8-17C}
\end{align}
holds for an arbitrary element $\bm{s} \in  \mathfrak{A}^0$, 
any general adaptive attack $\bm{\alpha} \in \mathfrak{A}^2$ satisfies 
\begin{align}
I(\bm{M};\bm{Z})_{\Phi,\bm{\alpha}} =0. \Label{H8-17D}
\end{align}
\hfill $\square$\end{corollary}

This corollary guarantees that 
perfect security for any deterministic attack \eqref{H8-17C} implies
perfect security for any general adaptive attack \eqref{H8-17D}
without the linearity condition.
Notice that the mutual information leaked to wiretapper is not zero
in the counter example given  in Section \ref{S5}.

In the case of linear network codes, we have the following lemma,
which will be shown in the next subsection.

\begin{lemma}\Label{L5-10}
Let $M$ be the message and $L$ be the scramble random variable.
We assume that they are subject to the independent uniform distribution on ${\cal M}\times {\cal L}$.
For a  linear function $f_1$ from ${\cal M}\times {\cal L} \to {\cal M}$, 
we define the variable $X:=f_1(M,L)$ on ${\cal M}$.
We choose a linear function $g=(g_1,g_2)$ from 
${\cal M}\to {\cal M}\times {\cal L}$
such that $g(x) \in  f_1^{-1}(x)$, i.e., $ f_1 (g(x))=x$.
Then, 
\begin{align}
P_{M,X}(m,x)=P_{M,X}(m-g_1(x),0)\Label{5-10}.
\end{align}
\hfill $\square$\end{lemma}

When the message $M$ and the scramble random variable $L$
are subject to the independent uniform distribution,
applying Lemma \ref{L5-10} to the case when $X=(Y_{s_1}, \ldots, Y_{s_\zeta})$, 
we have 
\begin{align}
&H(\bm{M}| Y_{s_1}=z_{1}, \ldots, Y_{s_\zeta}=z_{\zeta})
\nonumber \\
=&
H(\bm{M}| Y_{s_1}=0, \ldots, Y_{s_\zeta}=0),
\end{align}
which implies the condition \eqref{H8-17B}.
Hence, Theorem \ref{T7B} guarantees the following theorem.

\begin{theorem}\Label{T7}
Assume that a network code $\Phi$ is linear with respect to a certain finite field $\FF_{q}$.
When the message $M$ and the scramble random variable $L$
are subject to the independent uniform distribution,
any general adaptive attack $\bm{\alpha} \in \mathfrak{A}^2$ satisfies 
\eqref{H8-17}.
\hfill $\square$\end{theorem}

Further, we have the following proposition.
\begin{proposition}[\protect{\cite[Theorem 1]{HOKC}}]\Label{T8}
Assume that a network code $\Phi$ is linear.
Any adaptive and active attack $(\bm{\alpha},\bm{\beta}) \in \mathfrak{A}^3$ satisfies 
\begin{align}
I(\bm{M};\bm{Z})_{\Phi,(\bm{\alpha},\bm{\beta})}= I(\bm{M};\bm{Z})_{\Phi,\bm{\alpha}}.
\end{align}
\hfill $\square$\end{proposition}
Although the paper \cite{HOKC} shows Proposition \ref{T8} only 
for an active attack,
the proof can be extended to an adaptive and active attack.
That is, the reduction from an adaptive and active attack 
$(\bm{\alpha},\bm{\beta}) \in \mathfrak{A}^3$
to an adaptive attack $\bm{\alpha} \in \mathfrak{A}^2$
can be shown in the same way as \cite[Theorem 1]{HOKC}.
Therefore, 
when $\Phi$ is a linear code,
combing the above fact and \eqref{ER},
we find the relations
\begin{align}
\max_{\bm{\alpha} \in \mathfrak{A}^3}
I(\bm{M};\bm{Z})_{\Phi,\bm{\beta}}
=&
\max_{(\bm{\alpha},\bm{\beta}) \in \mathfrak{A}^2}
I(\bm{M};\bm{Z})_{\Phi,\bm{\beta}} \nonumber \\
=&
\max_{\bm{s} \in \mathfrak{A}^0}
I(\bm{M};\bm{Z})_{\Phi,\bm{s}}.\Label{ER9}
\end{align}
That is, when a network code is linear,
we can restrict Eve's attacks to deterministic attacks.

\begin{remark}\rm
Here, we remark the difference between our adaptive attack and the adaptive attack in 
\cite{SHIOJI}.
The paper \cite{SHIOJI} considers the following attack when 
the code has block length $n$ and the sender sends information to the receiver $n$ times.
The eavesdropper can change the nodes to be attacked on the $i$-th transmission 
by using the information obtained by the previous attacks.
However, in our setting, 
the eavesdropper can change the node to be attacked 
during one transmission from the sender to the receiver.
\hfill $\square$\end{remark}

\begin{remark}\Label{R9}\rm
Here, we summarize the preceding studies \cite{KMU,Yao2014,Zhang,HOKC} for the security analysis on the active attack, which makes contamination of the information on the network.
The secrecy analysis in \cite{KMU,Yao2014,Zhang}
is different from the analysis in \cite{HOKC}
although the non-local code construction in \cite{KMU,Yao2014,Zhang} does not depend on the concrete form of matrices corresponding to the network topology, which is similar to our non-local code construction.

While the papers \cite{SK,Yao2014} considered correctness when the error exists,
it discusses the secrecy only when there is no error.
Indeed, the papers \cite{SK,Yao2014} provided a statement similar to the main result of the paper \cite{HOKC}.
However, while it showed the correctness under the presence of the contamination in a certain class, 
it showed only the secrecy without contamination.
However, the paper \cite{HOKC} showed the secrecy as well as  
the correctness under the presence of the contamination in a certain class.

While the papers \cite[Proposition 5]{KMU}\cite{Zhang} considered the secrecy when the error exists,
it addressed the amount of leaked information only when the eavesdropper does not know the information of the noise.
However, the analysis in \cite{HOKC}
evaluated the leaked information when the eavesdropper knows the information of the noise.

Further, the paper \cite{HOKC}
gave an example where the contamination improves her performance,
which was shown in Section \ref{S5}.
The code given in this example is imperfectly secure 
for the deterministic attack on any allowed pair of edges.
However, 
it is not unsecure for a certain active attack on the same allowed pair of edges.
\hfill $\square$\end{remark}

\begin{remark}\Label{R7}\rm
There is a possibility that the randomness given in each node
is not uniform.
In this case, it is usual to apply univeral2 hash function \cite{Carter,WC81,Hayashi-Tsurumaru}.
Then, the resultant variable is close to the uniform random variable.
That is, the variational distance between the distribution of the resultant variable
and the uniform distribution is upper bounded by $\epsilon$ \cite{bennett95privacy,HILL,hayashi11,H-tight}.
However, we cannot say that it is the uniform random variable.

To discuss such a case, we often employ another criterion, the variational distance criterion, 
in which, we focus on the variational distance $D_V$
between 
the joint distribution 
$P_{\bm{M},\bm{Z}}$
and 
the product distribution $P_{\bm{M}}\times P_{\bm{Z}}$
instead of the mutual information.
When the randomness given in each node is uniform,
we denote the resultant joint distribution 
and the mutual information 
by $P_{i,\bm{M},\bm{Z}}$ and
$I(\bm{M};\bm{Z})_{\Phi,(\bm{\alpha},\bm{\beta})}$.

When the randomness given in each node is given in the above case,
we denote the resultant joint distribution 
by $P_{r,\bm{M},\bm{Z}}$.
Since 
the variational distance between the distribution of the resultant variable
and the uniform distribution is upper bounded by $\epsilon$,
information processing inequality guarantees that
\begin{align}
D_V(P_{i,\bm{M},\bm{Z}}, P_{r,\bm{M},\bm{Z}})
\le \epsilon.
\end{align}
Pinsker inequality guarantees 
\begin{align}
D_V(P_{i,\bm{M},\bm{Z}},P_{\bm{M}}\times P_{\bm{Z}})
\le
\sqrt{\frac{1}{2}I(\bm{M};\bm{Z})_{\Phi,(\bm{\alpha},\bm{\beta})}}.
\end{align}
Therefore,
the secrecy in the above case is evaluated as
\begin{align}
& D_V(P_{r,\bm{M},\bm{Z}},P_{\bm{M}}\times P_{\bm{Z}})
\nonumber \\
\le &
D_V(P_{i,\bm{M},\bm{Z}}, P_{r,\bm{M},\bm{Z}})
+
D_V(P_{i,\bm{M},\bm{Z}},P_{\bm{M}}\times P_{\bm{Z}})
\nonumber \\
\le & \epsilon+
\sqrt{\frac{1}{2}I(\bm{M};\bm{Z})_{\Phi,(\bm{\alpha},\bm{\beta})}}.
\end{align}
\hfill $\square$\end{remark}
\subsection{Proofs of Theorem \ref{T7B} and Lemma \ref{L5-10}}

\begin{proofof}{Theorem \ref{T7B}}
We have
\begin{align}
& H(\bm{M}|\bm{Z})_{\Phi,\bm{\alpha}} \nonumber \\
= &
\sum_{z_1}
\sum_{z_2} 
\cdots
\sum_{z_{\zeta}} 
P_{Y_{\alpha_1}, Y_{2 (z_1)},\ldots, 
Y_{\alpha_\zeta (z_1,z_2, \ldots, z_{\zeta-1})}}
\nonumber \\
& \cdot
H(\bm{M}| Y_{\alpha_1}\!=\!z_{1}, 
Y_{\alpha_2(z_s1)}\!=\!z_{2}, \ldots, \!
Y_{\alpha_\zeta (z_1,z_2, \ldots, z_{\zeta-1})}\!=\!z_{\zeta})
\nonumber \\
= &
\sum_{z_1}
\sum_{z_2} 
\cdots
\sum_{z_{\zeta}} 
P_{Y_{\alpha_1}, Y_{2 (z_1)},\ldots, 
Y_{\alpha_\zeta (z_1,z_2, \ldots, z_{\zeta-1})}}
\nonumber \\
&\cdot H(\bm{M}| Y_{\alpha_1}, 
Y_{\alpha_2(z_s1)}, \ldots, 
Y_{\alpha_\zeta (z_1,z_2, \ldots, z_{\zeta-1})})
\nonumber \\
\ge &
\min_{ \bm{s} \in  \mathfrak{A}^0}
H(\bm{M}| Y_{s_1}, \ldots, Y_{s_\zeta}).
\end{align}
This relation implies \eqref{H8-17}.
\end{proofof}

\begin{proofof}{Lemma \ref{L5-10}}
Given $x,m$, we have
\begin{align}
\{ l | f_1(m,l)=x\}
=
\{ l | f_1(m- g_1(x),l-g_2(x))=0\}.
\end{align}
So, we have
\begin{align}
|\{ l | f_1(m,l)=x\}|
=
|\{ l | f_1(m- g_1(x),l)=0\}|.
\end{align}
Hence, we have \eqref{5-10}.
\end{proofof}

\section{Network with powerful adaptive attack}\Label{S5}
In this section, to consider when adaptive attack is more powerful than deterministic attack,
we address the single shot setting, in which, the sender sends only one element of $\FF_p$,
which is called the scalar linearity.
This section mainly addresses the scalar linearity although Theorem \ref{T7B} holds under vector linearity.

It is known that there exists a linear imperfectly secure code 
over a finite field $\FF_q$ of a sufficiently large prime power $q$
when Eve may access a subset of channels that does not contain a cut between Alice and Bob 
even when the linear code does not employ private randomness in the intermediate nodes \cite{Bhattad}\footnote{In contrast, the paper \cite{SK} discussed a similar code construction
by increasing $n$ (vector linearity) while it did not increase the size of $q$.
The paper \cite{CCMG} extended this type of vector linearity setting 
 of imperfectly secure codes to the case with multi-source multicast.}.
The rigorous definition of imperfectly secure code is given in the next paragraph.
Theorem \ref{T7B} guarantees that 
such a linear code is still imperfectly secure even for active and adaptive attack over the same network.
However, it is not clear whether there exists such a linear imperfectly secure code over a finite field $\FF_p$ of prime $p$.
The previous paper \cite[Section VII]{HOKC} showed that 
there exists no imperfectly secure code over active attacks under a toy network
while there exists an imperfectly secure code over deterministic attacks.
In that network model, non-linear code realizes the imperfect security over active attacks.
In this section, we show that there exists no imperfectly secure code over adaptive attacks
in the same network model.

The toy network model given in \cite[Section VII]{HOKC} is the network of Fig. \ref{F1}, whose edges are 
$E=\{e(1),e(2),e(3),e(4)\}$.
Each edge $e(i)$ is assumed to send the binary information $\vec{Y}_i$.
No scramble random variable is allowed in the intermediate node, 
which is the condition (C2).
Eve is allowed to attack two edges of $E$ except for the pairs $\{e(1),e(2)\}$ and $\{e(3),e(4)\}$.
That is, $\mathfrak{A}^0=
\{ \{e(1),e(3)\},
\{e(2),e(3)\},
\{e(1),e(4)\},
\{e(2),e(4)\}\}$.
We adopt an imperfect security criterion in this section.
When ${Z}_E$ is Eve's information and $I(M;{Z}_E)<\log p$ for all of Eve's possible attacks,
we say that the code is {\it imperfectly secure} \cite{Bhattad} (or {\it weakly secure}).
Otherwise, it is called insecure.
That is, when there exists no function $\tilde{\psi}$ such that 
$\tilde{\psi}({Z}_E)=M$, our code is imperfectly secure.
Also, 
when ${Z}_E$ is Eve's information and $I(M;{Z}_E)=0$ for all of Eve's possible attacks,
we say that the code is {\it perfectly secure}.

\begin{figure}[htbp]
\begin{center}
\includegraphics[scale=0.5]{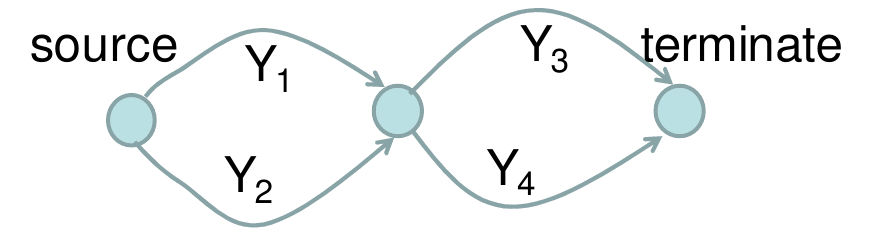}
\end{center}
\caption{Non-linear code.}
\Label{F1}
\end{figure}%

Let $L\in \FF_p$ be the uniform scramble random variable generated in the source node.
Assume that the intermediate node generates another uniform scramble random variable $L'\in \FF_p$.
The following scalar-linear code is perfectly secure.
The encoder $\phi$ is given as 
\begin{align}
{Y}_1:=L, \quad
{Y}_2:=M+L. \Label{Eq3B}
\end{align}
Then, the intermediate node makes the code $\varphi$ as
\begin{align}
{Y}_3:= L',\quad
{Y}_4:=Y_2-Y_1+L'. \Label{Eq2F}
\end{align}
The decoder $\psi$ is given as $\psi({Y}_3,{Y}_4):={Y}_4-{Y}_3$,
which equals $Y_2-Y_1+L'-L'=M+L-L +L'-L'=M$.
Then, the pair $(Y_1,Y_3)$ is independent of $M$.
Similarly, the pairs $(Y_1,Y_4)$, $(Y_2,Y_3)$, and $(Y_2,Y_4)$ are independent of $M$.
Hence, this code is perfectly secure for deterministic attack, and has the transmission rate $\log p$.
Due to the linearity, it is secure even for active and adaptive attack.
However, when the intermediate node cannot generate another 
uniform scramble random variable,
as shown in \cite[Theorem 4 of Section VII]{HOKC},
there is no imperfectly secure scalar-linear code over finite field $\FF_p$ with prime $p$ for deterministic attacks.
In other words, no scalar-linear code over finite field $\FF_p$ can realize the situation that
Eve cannot recover the message $M$ perfectly with deterministic attack.

To resolve this problem, there are two methods.
One is use of vector-linearity, and the other is use of non-linear code.
To use vector-linear code,
we consider the case when the network of Fig. \ref{F1} is used twice.
Assume that the source node generates three uniform scramble random variables $L_1,L_2,L_3\in \FF_p$.
Using these variable, we give a vector-linear code as follows.
The first transmission sends the following;
\begin{align}
{Y}_1:=L_1, \quad {Y}_2:=M+L_1, \Label{Eq3C}
\end{align}
and the second transmission sends the following;
\begin{align}
{Y}_1':=L_2, \quad {Y}_2':=L_3+L_2, \Label{Eq3D}
\end{align}
Then, the intermediate node makes the code $\varphi$ as
\begin{align}
{Y}_3:= Y_2'-Y_1',\quad
{Y}_4:=Y_2-Y_1+Y_2'-Y_1'. \Label{Eq2G}
\end{align}
In this code, nothing is transmitted in the second layer at the second transmission.
The decoder $\psi$ is given as $\psi({Y}_3,{Y}_4):={Y}_4-{Y}_3$,
which equals $Y_2-Y_1+Y_2'-Y_1'-(Y_2'-Y_1')=M+L_1-L_1=M$.
Then, the pair $(Y_1,Y_3)$ is independent of $M$.
Similarly, the pairs $(Y_1,Y_4)$, $(Y_2,Y_3)$, and $(Y_2,Y_4)$ are independent of $M$.
Hence, this code is perfectly secure for deterministic attack, and has the transmission rate $\frac{1}{2}\log p$.
Due to the linearity, it is secure even for active and adaptive attack.
In this code, 
the code in the second layer is composed of the message $Y_2-Y_1$
and the scramble $Y_2'-Y_1'$.
The secrecy of both are required in the transmission in the first layer.
Hence, totally three scramble variables are required in the source node.

As another solution, we discuss non-linear code as follows.
For this aim, we consider the case when the sender transmits only the binary message $M \in \FF_2$ and any edge can transmit only a binary information.
Now, we prepare the binary uniform scramble random variable $L\in \FF_2$.
We consider the following code.
The encoder $\phi$ is given in the same way as \eqref{Eq3B}.
Then, we consider non-linear code $\varphi$ in the intermediate node as
\begin{align}
{Y}_3&:= {Y}_1({Y}_2+{Y}_1)={Y}_1({Y}_2+1), \Label{Eq1}\\
{Y}_4&:=({Y}_1+1)({Y}_2+{Y}_1)
=({Y}_1+1){Y}_2. \Label{Eq2}
\end{align}
The decoder $\psi$ is given as $\psi({Y}_3,{Y}_4):={Y}_3+{Y}_4$.
Since ${Y}_3$ and ${Y}_4$ are given as follows under this code;
\begin{align}
{Y}_3= LM, \quad
{Y}_4=LM+M,
\end{align}
the decoder can recover $M$ nevertheless the value of $L$.

The leaked information for the deterministic attack is calculated as follows.
As shown in \cite[Appendix B]{HOKC},
the mutual information and the $l_1$ norm security measure
of these cases are calculated to 
\begin{align}
 & I(M;{Y}_1,{Y}_3)= I(M;{Y}_1,{Y}_4) \nonumber \\
  =& I(M;{Y}_2,{Y}_3)=  I(M;{Y}_2,{Y}_4)=\frac{1}{2},\Label{E9}\\
 & d_1(M|{Y}_1,{Y}_3)= d_1(M|{Y}_1,{Y}_4) \nonumber \\
  =& d_1(M|{Y}_2,{Y}_3)=  d_1(M|{Y}_2,{Y}_4)=\frac{1}{2},\Label{E10}
\end{align}
where the $l_1$ norm security measure $d_1(X|Y)$ is defined as
$d_1(X|Y):=\sum_y \sum_{x}
|\frac{1}{|{\cal X}|}P_Y(y) -P_{XY}(xy)|$
by using the cardinality $|{\cal X}|$ of the set of outcomes of the variable $X$.
In this section, we choose the base of the logarithm to be $2$.
Therefore, we find that this code is secure for deterministic attacks.
That is, we find that there exists a secure code over deterministic attacks.
Further, as shown in Proposition \ref{T6}, 
when Eve cannot recover the message $M$ perfectly with any deterministic attack in the code,
the network code is limited to this code or a code equivalent to this code.
This fact shows that there exists no imperfectly secure code over active attacks.

\begin{proposition} \cite[Lemma 4 of Section VII]{HOKC}\Label{T6}
Assume that a code $(\phi,\varphi,\psi)$ satisfies the following conditions.
Let $Y_1$ and $Y_2$ be the random variable generated by the encoder 
$\phi$ when $M$ is subject to the uniform distribution.
We assume that the random variables $(Y_3,Y_4):=\varphi (Y_1,Y_2)$
satisfies the following conditions.
\begin{description}
\item[(C1)] The relation $\psi (Y_3,Y_4)=M$ holds.
\item[(C2)] There is no deterministic function $\tilde{\psi}$
from $\FF_2^2$ to $\FF_2$ satisfying one of the following conditions.
\begin{align}
\tilde{\psi}(Y_1,Y_3)=M,\quad
\tilde{\psi}(Y_1,Y_4)=M, \\
\tilde{\psi}(Y_2,Y_3)=M,\quad
\tilde{\psi}(Y_2,Y_4)=M.
\end{align}
\end{description}
Then, there exist functions
$f_1,f_2,f_3,f_4$ on $\FF_2$ such that
$Y_i':=f_i(Y_i)$ is given in \eqref{Eq1}, \eqref{Eq2}, and \eqref{Eq3}
with a scramble random variable $L$ while the variable $L$ might be correlated with $M$.
\hfill $\square$\end{proposition}

Now, we show that
there exists no imperfectly secure code even for adaptive attacks without active modification.
Due to the above observation, it is sufficient to show that
there exists an adaptive attack to recover the message $M$
for the above given code.
Here, we give two types of adaptive attacks to recover the message $M$ as follows. 

\begin{description}
\item[(i)]
First, Eve eavesdrops $e(1)$.
When ${Y}_1=1$, 
she eavesdrops $e(3)$.
Then, she recovers $M$ as ${Y}_3={Y}_2+1= {Y}_2+{Y}_1=M$.
When ${Y}_1=0$, she eavesdrops $e(4)$
Then, she recovers $M$ as ${Y}_4={Y}_2={Y}_2+{Y}_1=M$.

\item[(ii)]
First, Eve eavesdrops $e(2)$.
When ${Y}_2=1$, 
she eavesdrops $e(4)$.
Then, she recovers $M$ as ${Y}_4={Y}_1+1= {Y}_1+{Y}_2=M$.
When ${Y}_2=0$, she eavesdrops $e(3)$
Then, she recovers $M$ as ${Y}_3={Y}_1={Y}_1+{Y}_2=M$.

\end{description}
Therefore, we find that
this code is not imperfectly secure even for adaptive attacks without active modification.
That is, there exists no imperfectly secure code over adaptive attacks
in this network model.
This fact shows that an adaptive attack is powerful for this kind of non-linear code as an active attack
even when it has no active modification.
The discussion in this section is summarized as Table \ref{non-linear}.

\begin{table}[htpb]
  \caption{Summary for one hop relay network (Fig. \ref{F1}) with single shot setting}
\Label{non-linear}
\begin{center}
  \begin{tabular}{|l|c|c|} 
\hline
\multirow{2}{*}{Code} 
& deterministic & adaptive \\
& attack & attack\\
\hline
\hline
scalar-linear code over $\FF_p$ & 
\multirow{2}{*}{insecure}& 
\multirow{2}{*}{insecure}\\
with prime $p$ & &\\
\hline
scalar-linear code over $\FF_q$  with &
imperfectly & 
imperfectly \\
sufficiently large prime power $q$ &secure&secure \\
\hline
\multirow{2}{*}{non-linear code over $\FF_2$} & imperfectly  & \multirow{2}{*}{insecure} \\
& secure& \\
\hline
\multirow{2}{*}{vector-linear code over $\FF_p$} & perfectly & perfectly \\
& secure & secure \\
\hline
  \end{tabular}
\end{center}
\end{table}

\section{Asymptotic formulation} \Label{S5B}
Next, given a network and the collection $\mathfrak{A}^0$,
we consider the capacity and the capacity region depending on 
the restrictions on the codes.
Due to \eqref{ER}, in the following, we do not consider randomization of 
Eve's attack.
We assume that each edge transmits $\{1, \ldots, d\}^n$
when we use channel at $n$ times, 
where the number $n$ is called the block-length.
Given integers $n$ and $d$,
we apply the formulation (including the linearity) given in Section \ref{S4-1}
to the case when ${\cal X}$ is given as $\{1,\ldots, d  \}^n$.
In this sense, the linearity condition (C1) is defined with block-length $n$,
and Theorem \ref{T7B} can be applied in this discussion.
Then, dependently of the block length $n$, 
we denote $\mathfrak{A}^i $ and $\mathfrak{C}^0$
by $\mathfrak{A}_n^i $ and ${\mathfrak{C}}_n^0 $, respectively,
although the collection $\mathfrak{A}_n^0$ does not depend on $n$.
First, we focus only on an adaptive attack $\bm{\alpha} \in \mathfrak{A}_n^2$.
Since there is no noise,
we denote the decoding error probability depends only on our code
$\Phi\in \mathfrak{C}_n^0$.
Hence, we denote it by $P_e(\Phi) $.
Then, we impose the following two conditions to our code 
$\Phi \in \mathfrak{C}_n^0$.
\begin{description}
\item[(C3)]
[Reliability]
The relation $P_e(\Phi) =0$.

\item[(C4)]
[Secrecy]
The relation $I(\bm{M};\bm{Z})_{\Phi,\bm{\alpha}}  =0$ holds for $\bm{\alpha} \in \mathfrak{A}_n^2$.
\end{description}
We denote the set of codes satisfying the above two conditions by $ \mathfrak{C}_n^1$.
Additionally, we denote the set of codes satisfying 
the no-randomness condition (C2) as well as these two conditions by $ \mathfrak{C}_n^2$.
In the unicast case, i.e., the case with $a=b=1$,
we define the full-randomness capacity $C_1$ 
and the no-randomness capacity $C_2$ as
\begin{align}
C_i& :=\sup_{n} \sup_{ \Phi \in \mathfrak{C}_n^i}
\frac{1}{n}\log | \Phi| ,\quad i=1,2.
\end{align}
Here, we should remark that we impose no linearity condition for our code.
From the definition, we have the relation
\begin{align}
C_2 \le C_1
\Label{E9-23}.
\end{align}

In the multiple multicast case, we define the full-randomness capacity region ${\cal C}_1$ 
and the no-randomness capacity region ${\cal C}_2$ as
\begin{align}
{\cal C}_{i'} :=
\sup_{n}\sup_{ \Phi \in \mathfrak{C}_n^{i'}}
\{(\frac{1}{n}\log |\Phi|_{i,j})_{i,j}\}
,\quad i'=1,2.
\end{align}

Similar to \eqref{E9-23}, we have the relation
\begin{align}
{\cal C}_2 \subset {\cal C}_1
\Label{E9-24}.
\end{align}

Next, we consider the case when each node has limited randomness, which is given as the condition (C2').
Since this generalized case is complicated, we discuss this generalized setting only with the unicast case.
Further, we suppose that each group is composed of one node.
Then, as in the condition (C2'),
we assume that the node in $i$-th group can use $\gamma_i$ random numbers $T_i$ per transmission.
We denote the set of codes satisfying this condition with length $n$
by $\mathfrak{C}_n[ (\gamma_i)_i]$.
Then, we define the capacity $C[ (\gamma_i)_i]$ 
with limited randomness as
\begin{align}
C[ (\gamma_i)_i] & :=\sup_{n} \sup_{ \Phi \in \mathfrak{C}_n[ (\gamma_i)_i ]}
\frac{1}{n}\log | \Phi| .
\end{align}

To clarify the effect by the linearity restriction,
we denote the capacity and capacity region 
by $C_{i,L}$ and ${\cal C}_{i,L}$, respectively
when the linearity restriction (C1) is imposed to our codes. 
Then, we have the relation $C_{i,L} \le C_{i}$ and 
${\cal C}_{i,L}\subset {\cal C}_{i}$.
Also, the capacity with limited randomness with linearity restriction (C1) to our codes
is denoted by $C[ (\gamma_i)_i]_L$.

Restricting Eve's attack to the deterministic attacks $ \mathfrak{A}_n^0$,
we define the above type of capacities and capacity regions, which are denoted by 
$C_{i,D}, C_{i,L,D},C[ (\gamma_i)_i]_{D}, C[ (\gamma_i)_i]_{L,D}, 
{\cal C}_{i,D}$ and ${\cal C}_{i,L,D}$, respectively. 
Then, we have the relations 
$C_{i,L,D} = C_{i,L}$,
$C_{i,D} \ge C_{i}$,
${\cal C}_{i,L,D}= {\cal C}_{i,L}$, 
${\cal C}_{i,D}\supset {\cal C}_{i}$, 
and the similar relations.

Now, we address the case when an adaptive and active attack $\bm{\beta} \in \mathfrak{A}_n^3$ is allowed
for Eve.
In this case, we replace the condition (C4) by the following condition;
\begin{description}
\item[(C4')]
[Secrecy]
The relation $I(\bm{M};\bm{Z})_{\Phi,\bm{\beta}}  =0$ holds for $\bm{\beta} \in \mathfrak{A}_n^3$.
\end{description}
However, we do not replace (C3) by the following 
robustness condition;
\begin{align}
P_e(\Phi,\bm{\beta}) =0 \hbox{ for } \forall 
\bm{\beta} \in \mathfrak{A}_n^3\Label{E5-10-A},
\end{align}
where $P_e(\Phi,\bm{\beta}) $ is the decoding error probability 
with our code $\Phi$ when Eve makes the attack $\bm{\beta}$. 
This situation can be justified in the following way when free public channel with no error is available.
In this case, to communicate each other securely, 
they need to share secret random variables.
To generate secret random variables,
they send secret random variables via the secure network coding.
The secrecy of the generated random variables is guaranteed by the secrecy condition (C4).
That is, condition (4) is definitely needed.
However, the robustness condition \eqref{E5-10-A}
is not necessary because they can check whether the transmitted random number is correct 
when the error verification test with the public channel is available after the transmission \cite[Section VIII]{Fung} \cite[Step 4 of Protocol 2]{H17} (See Remark \ref{R8}.).
Hence, we impose the condition (C3) instead of \eqref{E5-10-A}.
Replacing the condition (C4) by the condition (C4'), 
we define the above type of capacities and capacity regions, which are denoted by
$C_{i,AC}, C_{i,L,AC},C[ (\gamma_i)_i]_{AC}, C[ (\gamma_i)_i]_{L,AC}, 
{\cal C}_{i,AC}$ and ${\cal C}_{i,L,AC}$, respectively. 
Then, we have the relations 
$C_{i,L,AC} = C_{i,L}$,
$C_{i,D} \ge C_{i,AC}$,
${\cal C}_{i,L,AC}= {\cal C}_{i,L}$,
${\cal C}_{i,D}\supset {\cal C}_{i,AC}$, and the similar relations.
In summary, 
for each $i=1,2$, 
we have
\begin{align}
{\cal C}_{i,L,AC}={\cal C}_{i,L}={\cal C}_{i,L,D}\subset
{\cal C}_{i,AC}\subset {\cal C}_i\subset {\cal C}_{i,D}.\Label{LLP}
\end{align}
That is, when the equality ${\cal C}_{i,D}={\cal C}_{i,L,D}$ holds,
all the capacities have the same value. 
In other cases, we have similar relations.
 
\begin{remark}\Label{R8}\rm
When the public channel is available, 
the error verification can be done as follows.
Alice and Bob apply a universal2 hash function to their respective message with output length $m_2$.
They exchange their output of the universal2 hash function via the public channel.
If they are the same,
they consider that the message was transmitted correctly.
If they are different, they consider that 
the message was transmitted incorrectly.

As a typical example of a universal2 hash function,
we can use a modified Toeplitz matrix 
whose detail construction and evaluation of the complexity of its construction
are summarized in the recent paper \cite[Appendix]{Hayashi-Tsurumaru}.
Its calculation complexity is $O(m \log m)$ when $m$ is the input bit length.
Indeed, it was reported in paper \cite{Hayashi-Tsurumaru} that
the above type hash function practically implemented with $m=1 000 000$
by a conventional personal computer.

Due to this step, we can guarantee the correctness with probability $1-2^{-m_2}$, which is called the significance level\cite[Section VIII]{Fung}.
So, it is enough to choose $m_2$ depending on the required significance level.
This evaluation means that we do not need to increase the bit length $m_2$ for error verification
even when the length of message increases.

Here, one might care of the information leakage due to the information exchange on the public channel.
They can avoid such information leakage 
when they share $m_2$ bit common secret bits priorly. 
As above mentioned, the length $m_2$ of shared  
secret bits does not depend on the length of message.
Hence, when the length of message is very large,
the length $m_2$ of shared  
secret bits is negligible.
\hfill $\square$\end{remark}

\begin{example}\Label{ST}\rm
Now, as a typical example, we consider a single source acyclic network where Eve may choose any $r$-subset channels to access, which we call $r$-wiretap network \cite{Cai2002,CY,ACLY,LYC}.
That is, $\mathfrak{A}^0 $ is given as $\{ \bm{s} \subset [\ell] :| \bm{s}|=r\}$.
To discuss the capacities of the given network, we introduce two kinds of minimum cuts.
To define them,
we define a pseudo source node as a node that has only out-going edges but has no original message to be transmitted.
A pseudo source node is classified as an intermediate node because it is not the source node nor the terminal node.
The first type of minimum cut $\mincut_1$ is the minimum number of edges crossing a line separating
the source node and the terminal node.
The second type of minimum cut $\mincut_2$ is the minimum number of edges crossing a line separating
the source node and the terminal node with removing all edges out-going from pseudo source nodes.
That, while edges out-going from pseudo source nodes are ignored in $\mincut_2$,
they are counted in $\mincut_1$.
For $r$-wiretap network,
we have 
\begin{align}
&C_{2,L,AC}=
C_{2,L}=C_{2,L,D}=
C_{2,AC}= C_2= C_{2,D}\nonumber\\
=&
\mincut_2-r,\Label{E9-2} \\
& \mincut_2-r\le
C_{1,L,AC}=C_{1,L}=C_{1,L,D}\le
C_{1,AC} \nonumber \\
\le & C_1\le C_{1,D}\le 
\mincut_1-r.\Label{E9-2B}
\end{align}
When the network has no pseudo source node, 
$\mincut_2=\mincut_1$, which implies the equalities in \eqref{E9-2B}.
For example, the network given in Fig. \ref{5nodes} shows a network has different rates $\mincut_1$ and $\mincut_2$. 
This network has a linear code to realize $\mincut_1-r$ when $r=1$, which implies the equalities in \eqref{E9-2B}.

\begin{figure}[htbp]
\begin{center}
\includegraphics[scale=0.5]{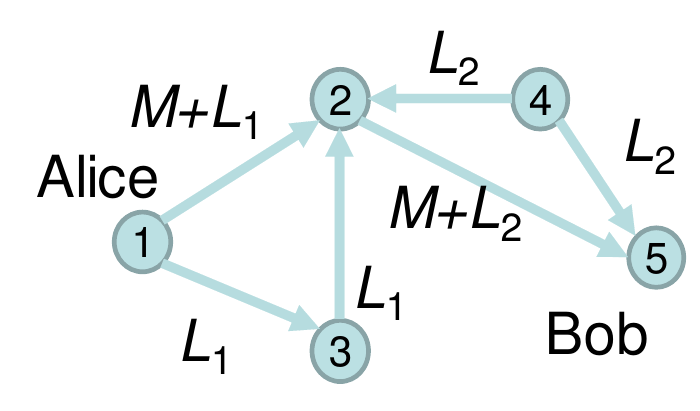}
\end{center}
\caption{Network with equality in \eqref{E9-2B}.
Node 1 is the source node and Node 5 is the terminal node.
Node 4 is a pseudo source node.
Hence, $\mincut_2=1$ and $\mincut_1=2$.
It also shows a linear code to achieve $\mincut_1-r$ when $r=1$.
The source node (Node 1) has the message $M$ and a scramble variable $L_1$.
The pseudo source node (Node 4) has another scramble variable $L_2$.
Even when Eve wiretaps any one edge, she cannot obtain any information for the message $M$.}
\Label{5nodes}
\end{figure}%

The relations \eqref{E9-2} and \eqref{E9-2B} can be shown as follows. 
It was shown in \cite[Section III]{CY} that 
the rate $\mincut_2-r$ 
is achievable by a linear code where only source node generates randomness
when Eve is allowed to use deterministic attack.
However, any adaptive and active attack is reduced to deterministic attack under a linear code.
Hence, we obtain $ C_{2,L,D} \ge \mincut_2-r$.

Using a idea similar to \cite[Section IV]{CY},
we show $C_{1,D} \le \mincut_1-r$. For this aim,
we choose edges crossing a line separating
the source node and the terminal node such that 
these edges contains the $r$ eavesdropped edges.
Let $Z$ be the variable on the $r$ eavesdropped edges, 
and $Y$ be the variable on the above edges crossing the separating line. 
Let $M$ be the message to be securely transmitted.
Due to the security condition, we have $I(M;Z)=0$
When an edge has an information with cardinality $d$,
the receiver's information $B$ satisfies
\begin{align}
& I(M;B)\le
I(M;Y)= I(M;YZ) \nonumber \\
=& I(M;Z)+I(M;Y|Z)
= I(M;Y|Z) \le H(Y|Z) \nonumber \\
\le & (\mincut_1-r) \log d ,
\end{align}
which implies
$C_{1,D} \le \mincut_1-r$.
Therefore, using \eqref{E9-23} and \eqref{LLP} and combining these facts, 
we obtain \eqref{E9-2B}.

When no intermediate node is allowed to generate randomness,
any pseudo source node plays no role.
Hence, the above discussion yields that
$C_{2,D} \le \mincut_2-r$.
Thus, we obtain \eqref{E9-2}.

\hfill $\square$\end{example}

\begin{example}\Label{Ex2}\rm
Next, we consider the case when $\mathfrak{A}^0$ is given by using the following group structure of the intermediate nodes.
The intermediate nodes are divided into $c-1$ groups, from the first group to the $c-1$-th group.
Here, $a$ source nodes and $b$ terminal nodes are regarded as
the $0$-th group and the $c$-th group, respectively.
For $i=1, \ldots, c$,
there are several edges between the $i-1$-th group and the $i$-th group.
We call the set of these edges the $i$-th edge group.
As seen later, this grouping of edges is essential to define the collection $\mathfrak{A}^0$.
Each intermediate node has incoming edges and outgoing edges.

Eve is assumed to eavesdrop a part of edges from the $i$-th edge group.
Eve's ability is characterized by the collection of subsets of the $i$-th edge group to be eavesdropped, which is called {\it the $i$-th tapped-edge collection} and is denoted by ${\cal S}_i$.
When an intermediate node of $i$-th group is directly linked to an intermediate node of $i+2$-th group,
we consider that the intermediate node of $i$-th group is connected 
to intermediate node of $i+1$-th group with an edge that is not contained in any member of 
the $i+1$-th tapped-edge collection ${\cal S}_{i+1}$.
Similarly, 
when an intermediate node of $i$-th group is directly linked to an intermediate node of $i+i'$-th group,
we can apply the same reduction.
Hence, without loss of generality, we can assume that
an outgoing edge of an intermediate node of $i$-th group is linked only to an intermediate node of $i+1$-th group.
Hence, the collection $\mathfrak{A}^0$
is given to be
${\cal S}_{1} \times {\cal S}_{2} \times \cdots \times {\cal S}_{c}$.

This example shows the following.
When the network model is composed of $c-1$ groups of intermediate nodes,
we can assume that the collection $\mathfrak{A}^0$
is given to be
${\cal S}_{1} \times {\cal S}_{2} \times \cdots \times {\cal S}_{c}$
without loss of generality.
\hfill $\square$\end{example}

\section{Relay network}\Label{S6}

\subsection{Formulation and capacities}
Now, as a special case of Example \ref{Ex2}, we consider the relay network given in Fig. \ref{FT} as a generalization of the network of Fig \ref{F1}.
This network is a unicast network, and only one intermediate node in each intermediate group.
That is, it has $c-1$ intermediate nodes.
We have $k_i$ edges between the $i-1$ and $i$-th nodes. 
In one channel use, each edge $e(i,j)$ can transmit the information ${Y}_{i,j}$ for $i=1,\ldots, c$ 
and $j=1, \ldots, k_i$ that takes values on $\{1, \ldots, d\}$.

\begin{figure}[htbp]
\begin{center}
\includegraphics[scale=0.5]{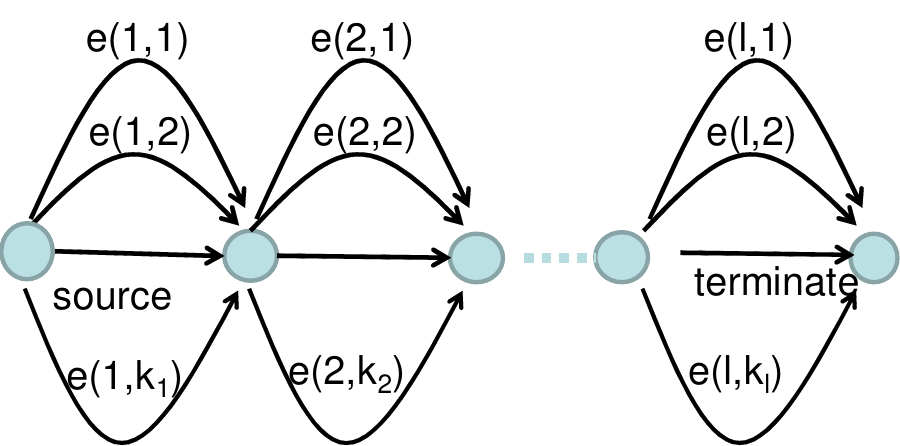}
\end{center}
\caption{Unicast relay network.}
\Label{FT}
\end{figure}%

Here, we assume that Eve can eavesdrop $r_i$ edges 
$\vec{Y}_{i,s_i}:=({Y}_{i, s_i(1)}, \ldots, {Y}_{i, s_i(r_i)}) $
among $k_i$ edges 
$\overline{Y}_{i}:=({Y}_{i, j})_{j=1, \ldots, k_i} $
between the $i-1$ and $i$-th nodes.
In this notation, the function $s_i$ expresses the edges eavesdropped by Eve.
That is, she can eavesdrop $\sum_{i=1}^c r_i$ edges totally.
In this paper, we allow stronger attacks for Eve than conventional attacks,
i.e., adaptive attacks and active attacks.

Then, we have the following capacity theorem.
\begin{theorem}\Label{TT3}
Defining
\begin{align}
h^1:=k_1 ,\quad
h^j:= \min( k_j ,
\frac{k_{j-1}-r_{j-1}}{k_{j-1}} h^{j-1}+\gamma_j 
),
\end{align}
we have
\begin{align}
C_1 =& 
C_{1,L} = 
C_{1,D} = 
C_{1,L,D} 
\nonumber \\
=& 
C_{1,AC} = 
C_{1,L,AC} = 
\log d \min_{1 \le j \le c} (k_j-r_j) \Label{Cap-1},
\\
C_2=& 
C_{2,L}= 
C_{2,D}= 
C_{2,L,D}
=
C_{2,AC}= 
C_{2,L,AC}
\nonumber \\
= &
\log d \min_{1 \le j \le c}
(k_j-r_j) 
\frac{(k_{j+1}-r_{j+1}) \cdots (k_c- r_c) }{k_{j+1} \cdots k_c },\Label{Cap-2}
\end{align}
and
\begin{align}
&C[ (\gamma_i)_i] =
C[ (\gamma_i)_i]_L =
C[ (\gamma_i)_i]_{D} =
C[ (\gamma_i)_i]_{L,D} 
\nonumber \\
=&
C[ (\gamma_i)_i]_{AC} 
=
C[ (\gamma_i)_i]_{L,AC} =
\log d \min_{1 \le j \le c}
\frac{k_j-r_j}{k_j}h^j .\Label{Cap-2B}
\end{align}
\hfill $\square$\end{theorem}

When the randomness is allowed in the intermediate nodes,
in the network of Fig \ref{F1}, the code given in \eqref{Eq3B} and \eqref{Eq2F} achieves the capacity 
$C_1$ as well as the capacities given in \eqref{Cap-1}.
In general,
the capacity $C_1$ is given as the rate of the bottleneck layer, which equals the minimum 
$\log d \min_{1 \le j \le c} (k_j-r_j) $.
When the randomness is not allowed in the intermediate nodes,
the code given in \eqref{Eq3C}, \eqref{Eq3D}, and \eqref{Eq2G} achieves the capacity 
$C_2$ as well as the capacities given in \eqref{Cap-2}.
In general,
the capacity $C_2$ is given as
the minium of the multiplication $\log d (k_j-r_j) 
\frac{(k_{j+1}-r_{j+1}) \cdots (k_c- r_c) }{k_{j+1} \cdots k_c }$ with respect to $j$.
Therefore, when $k_j$ and $r_j$ are constants $k$ and $r$,
the capacity $C_2$ is calculated to $\log d (k-r)^c/k^{c-1}$.
When $c$ goes to infinity, it converges to zero.

Here, we discuss the relation to existing results with respect to the difference between two capacities $C_1$
and $C_2$.
A larger part of existing studies discuss the capacity (or capacity region) with no restriction of
randomness generated in intermediate nodes.
For example, in $r$-wiretap network, which is a typical network model, as explained in Example \ref{ST}, 
the capacity with no restriction can be achieved without use of randomness generated in intermediate nodes.
However, the paper \cite{CY-07} showed an example, in which 
randomness generated in intermediate nodes improves the capacity.
In this example, the source node is connected only with one edge.
Usually, the secure transmission can be done by use of the difference between information on different edges connected to the same node.
Hence, it is natural that randomness generated in intermediate nodes improves the capacity
when each source node is connected only to one edge.

The papers \cite{CHK-10,CHK-13} addressed the difference between the existence and non-existence of 
randomness generated in intermediate nodes in another network
only for deterministic attacks.
However, they did not derive the capacities $C_{1,D}$ and $C_{2,D}$ exactly. Their analysis depends on special codes.
Therefore, our analysis is the first derivation of the difference between 
the capacities $C_{1,D}$ and $C_{2,D}$ except for the case when the source node is connected only with one edge.

\subsection{Converse part}
For any $j=1, \ldots, c$,
the rate of secure transmission from the $j-1$-th intermediate node to
the $j$-th intermediate node is $\log d (k_j-r_j) $.
Taking the minimum with respect to $j$,
we obtain $C_{1,D}\le \log d \min_{1 \le j \le c} (k_j-r_j)$.

Next, we consider \eqref{Cap-2}.
For the amount of leaked information, we have the following theorem.
\begin{theorem}\Label{T1}
Under the condition (C2), we have
\begin{align}
& \max_{s_1, \ldots, s_c} I(M; \vec{Y}_{1,s_1}, \ldots, 
\vec{Y}_{c,s_c}) \nonumber \\
\ge & H(M) 
\nonumber \\
&-
( \log d )
\min_{1 \le j \le c}
(k_j-r_j) 
\frac{(k_{j+1}-r_{j+1}) \cdots (k_c- r_c) }{k_{j+1} \cdots k_c }.
\end{align} 
\hfill $\square$\end{theorem} 

Therefore, to realize the condition 
\begin{align}
\max_{s_1, \ldots, s_c} I(M; \vec{Y}_{1,s_1}, \ldots, 
\vec{Y}_{c,s_c}) =0,
\Label{Hd1}
\end{align}
the message $M$ needs to satisfy the condition
\begin{align}
H(M)\le \log d
\min_{1 \le j \le c}
(k_j-r_j) 
\frac{(k_{j+1}-r_{j+1}) \cdots (k_c- r_c) }{k_{j+1} \cdots k_c }.
\Label{Hd3}
\end{align}
When use the same network $n$ times,
the condition \eqref{Hd1} requires the condition
\begin{align}
H(M)\le n \log d 
\min_{1 \le j \le c}
(k_j-r_j) 
\frac{(k_{j+1}-r_{j+1}) \cdots (k_c- r_c) }{k_{j+1} \cdots k_c },\Label{Hd2}
\end{align}
which implies $$C_{2,D}\le  
\log d \min_{1 \le j \le c}
(k_j-r_j) 
\frac{(k_{j+1}-r_{j+1}) \cdots (k_c- r_c) }{k_{j+1} \cdots k_c }.$$

Theorem \ref{T1} can be generalized to the limited randomness case as follows.
Hence, it is sufficient to show Theorem \ref{T1B}.
\begin{theorem}\Label{T1B}
Under the condition (C2'), we have
\begin{align}
& \max_{s_1, \ldots, s_c} I(M; \vec{Y}_{1,s_1}, \ldots, 
\vec{Y}_{c,s_c}) \nonumber \\
\ge & H(M) -
\log d \min_{1 \le j \le c}
\frac{k_j-r_j}{k_j}h^j .
\end{align} 
\hfill $\square$\end{theorem} 

\begin{proofof}{Theorem \ref{T1B}}
Now, we independently choose the sets $S_1, S_2, \ldots, S_c$
subject to the uniform distribution.
We denote the expectation is with respect to this random choice by $\mathbb{E}$.
We prove Theorem \ref{T1B} by using Lemma \ref{LH10}, which will be shown in the latter section.
Application of Lemma \ref{LH10} to $X=(\vec{Y}_{j-1,S_{j-1}}, \ldots, \vec{Y}_{2,S_{2}},\vec{Y}_{1,S_1})$
shows the inequality
\begin{align}
& \mathbb{E}
H(\vec{Y}_{j,S_j}|\vec{Y}_{j-1,S_{j-1}}, \ldots, \vec{Y}_{2,S_{2}},\vec{Y}_{1,S_1}) 
\nonumber \\
\ge &
\frac{r_j}{k_j}
\mathbb{E}H(\overline{Y}_{j}|\vec{Y}_{j-1,S_{j-1}}, \ldots, \vec{Y}_{2,S_{2}},\vec{Y}_{1,S_1})
\Label{eq-02}
\end{align}
for $1 \le j \le c$.
Then we have for $1 \le j \le c$,
\begin{align}
&\mathbb{E}H(M|\vec{Y}_{c,S_c},\vec{Y}_{c-1, S_{c-1}} \ldots, \vec{Y}_{2,S_2},\vec{Y}_{1,S_1}) \nonumber \\
 \le & \mathbb{E}H(M|\vec{Y}_{j,S_j},\vec{Y}_{j-1,S_{j-1}}, \ldots, \vec{Y}_{2,S_{2}},\vec{Y}_{1,S_1})\nonumber \\
\stackrel{(a)}{\le}  & \mathbb{E}H(\overline{Y}_{j} |\vec{Y}_{j,S_j},\vec{Y}_{j-1,S_{j-1}}, \ldots, \vec{Y}_{2,S_{2}},\vec{Y}_{1,S_1})\nonumber \\
=  & \mathbb{E} H( \vec{Y}_{j,S_j^c} |\vec{Y}_{j,S_j},\vec{Y}_{j-1,S_{j-1}}, \ldots, \vec{Y}_{2,S_{2}},\vec{Y}_{1,S_1})
\nonumber \\
=  & \mathbb{E} H( \overline{Y}_{j} | \vec{Y}_{j-1,S_{j-1}}, \ldots, \vec{Y}_{2,S_{2}},\vec{Y}_{1,S_1})
\nonumber \\
&- \mathbb{E} H( \vec{Y}_{j,S_j} | \vec{Y}_{j-1,S_{j-1}}, \ldots, \vec{Y}_{2,S_{2}},\vec{Y}_{1,S_1})
\nonumber \\
\stackrel{(b)}{\le} & \frac{k_j-r_j}{k_j}\mathbb{E} H(\overline{Y}_{j}|\vec{Y}_{j-1,S_{j-1}}, \ldots, \vec{Y}_{2,S_{2}},\vec{Y}_{1,S_1}),
\Label{eq-03}
\end{align}
where
$(a)$ follows from the fact that
$M$ is determined by the random variable $\overline{Y}_{j} $,
and $(b)$ follows from \eqref{eq-02}.

Similarly, we  have 
\begin{align}\label{eq-04}
&  \mathbb{E}H(\overline{Y}_{j}|\vec{Y}_{j-1,S_{j-1}}, \vec{Y}_{j-2,S_{j-2}},\ldots, \vec{Y}_{2,S_{2}},\vec{Y}_{1,S_1})\nonumber \\
\stackrel{(a)}{\le} & 
 \mathbb{E}H(\overline{Y}_{j-1},K_{j}|\vec{Y}_{j-1,S_{j-1}}, \vec{Y}_{j-2,S_{j-2}},\ldots, \vec{Y}_{2,S_{2}},\vec{Y}_{1,S_1})\nonumber \\
= & 
 \mathbb{E}H(\vec{Y}_{j-1,S_{j-1}^c},K_{j}|\vec{Y}_{j-1,S_{j-1}}, \vec{Y}_{j-2,S_{j-2}},\ldots, \vec{Y}_{2,S_{2}},\vec{Y}_{1,S_1})\nonumber \\
\le & 
 \mathbb{E}H(\vec{Y}_{j-1,S_{j-1}^c}|\vec{Y}_{j-1,S_{j-1}}, \vec{Y}_{j-2,S_{j-2}},\ldots, \vec{Y}_{2,S_{2}},\vec{Y}_{1,S_1})
\nonumber \\
& +\gamma_{j}\log d
\nonumber \\
= & 
 \mathbb{E}H(\overline{Y}_{j-1}| \vec{Y}_{j-2,S_{j-2}},\ldots, \vec{Y}_{2,S_{2}},\vec{Y}_{1,S_1})
\nonumber \\
& -\mathbb{E}H(\vec{Y}_{j-1,S_{j-1}}| \vec{Y}_{j-2,S_{j-2}},\ldots, \vec{Y}_{2,S_{2}},\vec{Y}_{1,S_1})
\nonumber \\
&  +\gamma_{j}\log d
\nonumber \\
\stackrel{(b)}{\le} & 
\frac{k_{j-1}-r_{j-1}}{k_{j-1}}\mathbb{E}H(\overline{Y}_{j-1}|\vec{Y}_{j-2,S_{j-2}}, \ldots, \vec{Y}_{2,S_{2}},\vec{Y}_{1,S_1})
\nonumber \\
& +\gamma_{j}\log d,
\end{align}
where 
$(a)$ follows from the fact that
$\overline{Y}_{j}$ is determined by the random variables $\overline{Y}_{j-1},K_{j}$,
and 
$(b)$ follows from \eqref{eq-02}.

Now, we show
\begin{equation}\label{eq-05}
\mathbb{E}H(\overline{Y}_{j}|\vec{Y}_{j-1,S_{j-1}}, \ldots, \vec{Y}_{2,S_{2}},\vec{Y}_{1,S_1}) \le  h^j  \log d
\end{equation}
by induction with respect to $j$.
Since $H(\overline{Y}_{1})\le k_1 $, \eqref{eq-05} holds for $j=1$.
Assume that
\begin{equation}
\mathbb{E}H(\overline{Y}_{j-1}|\vec{Y}_{j-2,S_{j-2}}, \ldots, \vec{Y}_{2,S_{2}},\vec{Y}_{1,S_1}) \le  h^{j-1}  \log d. \nonumber
\end{equation}
Then (\ref{eq-04}) implies that
\begin{align}
&\mathbb{E}H(\overline{Y}_{j}|\vec{Y}_{j-1,S_{j-1}}, \ldots, \vec{Y}_{2,S_{2}},\vec{Y}_{1,S_1})\nonumber \\
=&\frac{k_{j-1}-r_{j-1}}{k_{j-1}}
\mathbb{E}H(\overline{Y}_{j-1}|\vec{Y}_{j-2,S_{j-2}}, \ldots, \vec{Y}_{2,S_{2}},\vec{Y}_{1,S_1})
\nonumber \\
&+ \gamma_j \log d \nonumber \\
\le &  \frac{k_{j-1}-r_{j-1}}{k_{j-1}} h^{j-1} \log d+  \gamma_j\log d, \Label{Eq3}
\end{align}
Also, we have
\begin{align}
\mathbb{E}H(\overline{Y}_{j}|\vec{Y}_{j-1,S_{j-1}}, \ldots, \vec{Y}_{2,S_{2}},\vec{Y}_{1,S_1}) 
\le  H(\overline{Y}_{j}) \le k_{j}\log d. \label{Eq4}
\end{align}
Combining \eqref{Eq3} and \eqref{Eq4}, we have \eqref{eq-05}.

Therefore, combining \eqref{eq-03} and \eqref{eq-05},
we have
\begin{align}\label{eq-lm11B}
&\mathbb{E}H(M|\vec{Y}_{c,S_c},\vec{Y}_{c-1, S_{c-1}} \ldots, \vec{Y}_{2,S_2},\vec{Y}_{1,S_1}) 
\nonumber \\
\le &
 \frac{k_j-r_j}{k_j}  h^j \log d,
\end{align}
which is equivalent to
\begin{align}
& \mathbb{E}I(M;\vec{Y}_{c,S_c},\vec{Y}_{c-1, S_{c-1}} \ldots, \vec{Y}_{2,S_2},\vec{Y}_{1,S_1}) \nonumber \\
\ge &
H(M)- \frac{k_j-r_j}{k_j}  h^j \log d.
\label{eq-lm11}
\end{align}
Hence, we obtain the desired statement.
\end{proofof}

\subsection{Code construction to achieve capacity $C_{1,L,D}$}\Label{SS1}

We give a code to achieve the capacity $C_{1,L,D}$.
The idea of our construction is the same as a Wiretap-II code
introduced by Ozarow and Wyner \cite{Ozarow}. 
In wiretap channel II, 
a secure message is encoded to a codeword of $n$ length and wiretapper may access any $r$ components out of the $n$ components. 
Then, this secure code for wiretap channel II is secure 
even for our setting.
However, this construction will be applied to a more general case 
with modification in the latter section.
To discuss such a more general case, we need to concretely describe 
our whole construction to keep the self-consistency in this section.
For simplicity, we assume that the integer $d$ is a power $q$ of a prime $p$.
The general case will be discussed later.
When we can make the desired code in the case with $c=1$,
we can employ the constructed code for the secure transmission code from 
the $i-1$-th intermediate node to the $i$-th intermediate node
because the $i-1$-th intermediate node can employ scramble random numbers $T_{i-1}$.
For this purpose, we prepare the following lemma.

\begin{lemma}\Label{L425}
For any prime power $q$, any two natural numbers $k>r$,
there exist a natural integer $n_{k,r}$ 
and $r$ vectors $v_1, \ldots, v_r \in \FF_{q^{n_{k,r}}}^{k}$
such that $v_{i,j}=\delta_{i,j} $ for $j=1, \ldots,m$ and 
the $r \times r$ matrix $(v_{i,s(j)})_{i,j}$ is invertible
for any injective function $s$ from $\{1, \ldots, r\}$ to 
$\{1, \ldots, k\}$.
\hfill $\square$\end{lemma}

This lemma might be shown in the context of the wiretap channel II introduced by Ozarow and Wyner \cite{Ozarow}. 
In the model of wiretap channel II,
a secrete message is encoded to a codeword in an $n_{k,r}$-length code. 
A wiretapper may take any $r$ components 
out of $k$ parallel channels but may have no information about the message. 
A linear code, e.g., a Reed-Solomon code can serve as the code. 
It is called a $(k,r)$ code for wiretap channel II,
and satisfies the condition for Lemma \ref{L425}.
Also, this leamma also can be regarded as a very simple and special case of the code in \cite[Section III]{CY}.
For readers' convenience, we give its proof in Appendix \ref{A1}.

Here, we make the desired code in the case with $c=1$.
We employ the finite filed $ \FF_{q'}$ with $q'=q^{n_{k,r}}$.
That is, we need finite field of large size, whose efficient construction is discussed in \cite[Appendix D]{Hayashi-Tsurumaru}.
So, when we use the channel $n:=n' n_{k,r}$ times,
our transmission can be regarded as $n'$ times transmission on $ \FF_{q'} $,
i.e., each edge can transmit up to $n'$ symbols in $ \FF_{q'} $.
In the following, all random variables are treated as random variables taking values in $ \FF_{q'} $.

According to Lemma \ref{L425},
we choose $r$ vectors $v_1, \ldots, v_r \in \FF_{q^{n_{k,r}}}^{k}$.
Using $r $ additional scramble random numbers $L_1,\ldots, L_{r}$, 
we can transmit $k-r$ random variables $M_1,\ldots, M_{k-r} $ 
by encoding the random variable $\vec{Y}_j$ for the $j$-th edge by 
\begin{align}
{Y}_{j}:= 
\left\{
\begin{array}{ll}
L_j 
& \hbox{ when } j \le r \\
M_{j-r}+ \sum_{j'=1}^{r}v_{j',j}L_{j'} 
& \hbox{ when } r+1 < j \le k. 
\end{array}
\right.
\end{align}
Then, Bob recovers the original messages $M_1,\ldots, M_{k-r} $ as 
\begin{align}
M_j':= {Y}_{j+r}- \sum_{j'=1}^r v_{j',j} {Y}_{j'}.
\end{align}

Assume that Eve eavesdrops $r$ edges, the $s(1)$-th edge, $\ldots$, the $s(r)$-th edge.
Due to the condition in Lemma \ref{L425}, for any function $s$, 
the vectors $(v_{j',s(1)})_{1\le j'\le r}, \ldots, (v_{j',s(r)})_{1\le j'\le r}$
are linearly independent.
So, $\sum_{j'=1}^{r} v_{j',s(1)}L_{j'},$\par\noindent
$ \ldots, \sum_{j'=1}^{r} v_{j',s(r)}L_{j'}$
are $r$ uniform random numbers 
even when we fixed the values of the random variables $M_1,\ldots, M_{k-r} $.
Eve cannot obtain any information for  $M_1,\ldots, M_{k-r}$.

Repeating $n'$ times this procedure,
we can extend this method to the case when we transmit 
$(k-r)n'$ random variables $M_1,\ldots, M_{(k-r)n'} $ with
$r n'$ additional scramble random numbers $L_1,\ldots, L_{r n'} $.
Therefore, the transmission rate of this code is
$\frac{(k-r)\log_2 q'}{n_{k,r}}=(k-r)\log q $.
Since  $(M_1,\ldots, M_{(k-r)n'}) $ can be regarded as 
an element of a vector space over $\FF_{q'}$,
this operation is a linear code with respect to the finite field $\FF_{q'}$.
Therefore, since it satisfies the linearity condition (C1),
the above security analysis over the deterministic attack
guarantees the security over the adaptive and active attack due to Theorem \ref{T7}\footnote{
Theorem \ref{T7} can be applied to a linear code with respect to any finite field.
Hence, we do not need to restrict our discussion to linear codes with respect to the finite field $\FF_q$.}.

Here, we make the desired code in the case with general $c$.
Based on Lemma \ref{L425} with respective $k_i$ and $r_i$, we choose $n_{k_i,r_i}$.
Then, we choose the finite filed $ \FF_{q'}$ with $q'=q^{\overline{n}}$, where 
$\overline{n}:=\max_{1 \le i \le c}n_{k_i,r_i}$.
So, when we use the channel $n:=n' \overline{n}$ times,
our transmission can be regarded as $n'$ times transmission on $ \FF_{q'} $.
Therefore, we can transmit the minimum rate 
$ \log q \min_{1 \le j \le c} (k_j-r_j)$.
In this construction, the transmission on each step is given by a linear code over the finite field $\FF_{q'}$,
the whole operation is also given as a linear code over the finite field $\FF_{q'}$.
Therefore, since it satisfies the linearity condition (C1),
Theorem \ref{T7} guarantees the security over the adaptive and active attack.

The calculation complexity of this code can be evaluated as follows.
Node operations on node $j$ are written as $k_j \times k_j$ matrix multiplications over $ \FF_{q'}$.
When we choose a suitable algebraic extension $ \FF_{q'}$,
the multiplication over $ \FF_{q'}$ has complexity $O(\overline{n} \log  \overline{n})$.
Therefore, node operations on node $j$ has complexity $O(n' k_j^2 \overline{n} \log \overline{n})=O(n k_j^2 \log \overline{n})$.

Now, we consider the case that the integer $d$ is not a power $q$ of a prime $p$.
In this case, we have the following lemma.
\begin{lemma}\Label{LH3}
We have 
\begin{align}
\lim_{n \to \infty}
\frac{1}{n}
\log \max_{q: \hbox{ prime power}}
\{\log q | q \le d^n\}
= \log d.
\end{align}
\hfill $\square$\end{lemma}

Given a sufficiently large integer $n$,
we choose a prime power $q:=\argmax_{q: \hbox{ prime power}}\{\log q | q \le d^n\}$.
We treat $n$ uses of a channel as a single transmission of random variable taking values in $\FF_{q}$.
Due to Lemma \ref{LH3},
the code given above achieves the transmission rates
$ \log d \min_{1 \le j \le c} (k_j-r_j)$ 
when $n$ goes to infinity.

\subsection{Code construction to achieve capacity $C[(\gamma_i)_i]_{L,D}$ and $C_{2,L,D}$}\Label{S42B}
Since the capacity $C_{2,L,D}$ is a special case of $C[(\gamma_i)_i]_{L,D}$ with $\gamma_i=0$,
we construct only a code to achieve the capacity $C[(\gamma_i)_i]_{L,D}$.
Similar to the previous section, we choose the finite filed $ \FF_{q'}$ with 
$q'=q^{\overline{n}}$ and $\overline{n}:=\max_{1 \le i \le c}n_{k_i,r_i}$,
and
we consider the case of $n:=n' \overline{n}$ uses of the channel, i.e., 
each edge can transmit up to $n'$ symbols in $ \FF_{q'} $.
In the following, all random variables are treated as random variables taking values in $ \FF_{q'} $.
For notational simplicity, we consider the case when
single use of each edge transmits an element of $\FF_{q'}$.

To achieve the above purpose, we give a linear code with respect to $ \FF_{q'} $ satisfying the following two conditions (D1) and (D2) by induction with respect to $j$.
Since the code satisfies the linearity condition (C1),
it is sufficient to consider the deterministic attack.

\begin{description}
\item[(D1)]
The code securely transmits the message $M$ of $\underline{h}^j$
symbols per single use of channel 
to the $j$-th node from the source node,
where $\underline{h}^j:=\min_{1\le j' \le j} \frac{k_{j'}-r_{j'}}{k_{j'}}h^{j'} $.
That is, $ I(M; \vec{Y}_{1,s_1}, \ldots, 
\vec{Y}_{j,s_j}) =0$ for any $(s_1, \ldots, s_j)\in {\cal S}_1\times \cdots \times{\cal S}_j $.

\item[(D2)]
The $j$-th node receives secure random number $T_j'$ of $\overline{h}^j-\underline{h}^j$
symbols per single use of channel, which contains
the random numbers generated from the 1st node to the $j-1$-th node,
where $\overline{h}^j:=\frac{k_j-r_j}{k_j}h^j $.
That is,
the $j$-th node receives secure random number of $\overline{h}^j$
symbols per single use of channel, i.e.,
$ I(M T_{j}'; \vec{Y}_{1,s_1}, \ldots, 
\vec{Y}_{j,s_j}) =0$ for any $(s_1, \ldots, s_j)\in {\cal S}_1\times \cdots \times{\cal S}_j $.

\end{description}

Since $\overline{h}^1= \underline{h}^1= (k_1-r_1) $,
the desired code with $j=1$ was constructed in Subsection \ref{SS1}.
We show the existence of the desired linear code with respect to $ \FF_{q'} $
by induction. 
That is, we assume the existence in the case of $j-1$ with block length $n_{j-1}$.
We find that 
$\overline{h}^j  =\min
( (k_j-r_j) , \frac{k_j-r_j}{k_j} ( \overline{h}^{j-1} +\gamma_j ))$ 
and
$\underline{h}^j  =\min
( (k_j-r_j), \frac{k_j-r_j}{k_j} ( \overline{h}^{j-1} +\gamma_j ),
\underline{h}^{j-1} )$ 
for $j \ge 2$.
We show the existence of such a code with $j$ 
by classifying three cases.

(1) Case of $\underline{h}^j=\overline{h}^j=(k_j-r_j) $:
To achieve the desired task,
the $j-1$-th node needs to securely transmit the message $M$ of 
$(k_j-r_j) $ symbols per single use of channel
to the $j$-th node,
which requires scramble random numbers $T_j'$ of
$ {r_j} $ symbols per single use of channel
at the $j-1$-th node.
Since 
$ {r_j} \le \overline{h}^{j-1} +\gamma_j - (k_j-r_j) $,
the $j-1$-th node has sufficient 
scramble random numbers for this purpose.
We divide the scramble random numbers $T_j'$
into two parts $T_{j,1}'$ and $T_{j,2}'$, where
$T_{j,1}'$ has $\gamma_j $ symbols per single of channel 
and 
$T_{j,2}'$ has $({r_j} -\gamma_j )$ symbols per single of channel. 
Due to the assumption of induction,
the sender securely transmits $M $ and $T_{j,2}'$
to the $j-1$-th node by a linear code with block length $n'$,
where 
the first $n' (k_j-r_j) $ symbols are  $M $,
the next $n' (r_j -\gamma_j ) $ symbols are  $T_{j,2}'$,
and the remaining symbols are fixed to zero.
That is,
$ I(M T_{j,2}'; \vec{Y}_{1,s_1}, \ldots, 
\vec{Y}_{j-1,s_{j-1}}) =0$ for any $(s_1, \ldots, s_{j-1})\in {\cal S}_1\times \cdots \times{\cal S}_{j-1} $.
Since $ T_{j,1}'$ is composed of $n_{j-1} \gamma_j  $ symbols
and is independent of other random variables,
we apply the code given in Subsection \ref{SS1} 
to the message $M$ and the scramble random number $T_j'$.
Then, the $j-1$-th node securely transmits  
the message $M$ to the $j$-th node by a desired linear code with respect to $\FF_{q'}$ of  block length ${n}'$.
Therefore,
$ I(M; \vec{Y}_{j,s_j}|\vec{Y}_{1,s_1}, \ldots, 
\vec{Y}_{j-1,s_{j-1}}) =0$ for any $(j_1, \ldots, s_j)\in {\cal S}_1\times \cdots \times{\cal S}_j $.
Hence,
$ I(M; \vec{Y}_{1,s_1}, \ldots, 
\vec{Y}_{j,s_j}) =0$ for any $(j_1, \ldots, s_j)\in {\cal S}_1\times \cdots \times{\cal S}_j $.

(2) Case of $\underline{h}^j=\overline{h}^j=\frac{k_j-r_j}{k_j} ( \overline{h}^{j-1} +\gamma_j )$:
To achieve the desired task,
the $j-1$-th node needs to securely transmit the message $M$ of 
$\underline{h}^j=\overline{h}^j$ symbols per single use of channel
to the $j$-th node,
which requires scramble random numbers $T_j'$ of $\frac{k_j-r_j}{k_j} \overline{h}^j$ 
symbols per single use of channel
at the $j-1$-th node.
Since $\frac{k_j-r_j}{k_j} \overline{h}^j= \frac{r_j}{k_j-r_j}\frac{k_j-r_j}{k_j} ( \overline{h}^{j-1} +\gamma_j )  
=\overline{h}^{j-1} +\gamma_j - 
\frac{k_j-r_j}{k_j} ( \overline{h}^{j-1} +\gamma_j ) $,
the $j-1$-th node has sufficient scramble random numbers 
for the above purpose.
Therefore, similar to the case (1),
we can show the existence of the desired linear code with respect to
$\FF_{q'}$.

(3) Case of $\underline{h}^j= \underline{h}^{j-1}<\overline{h}^j$:
Since $\overline{h}^j$ is $k_j-r_j$ or $\frac{k_j-r_j}{k_j} ( \overline{h}^{j-1} +\gamma_j ) $,
due to the discussion with the above two cases (1) and (2),
the $j$-th node receives secure random number of $\overline{h}^j$
symbols per single use of channel.
To achieve the desired task,
the $j-1$-th node needs to securely transmit the message $M$ of 
$\underline{h}^j (\le k_j-r_j)$ symbols per single use of channel
to the $j$-th node,
which requires scramble random numbers $T_j'$ of $\frac{k_j-r_j}{k_j} \underline{h}^j$ 
symbols per single use of channel
at the $j-1$-th node.
Since $ \frac{r_j}{k_j-r_j}\underline{h}^{j}=\frac{r_j}{k_j-r_j}\underline{h}^{j-1}\le \gamma_j $,
the $j-1$-th node has sufficient scramble random numbers 
for this purpose.
Therefore, similar to the case (1),
we can show the existence of the desired linear code with respect to
$\FF_{q'}$.

Therefore, there exists a code that transmits the message with the rate $\underline{h}^c$
to the source node from the source node.
Due to the same discussion as Section \ref{SS1}
node operations on node $j$ has complexity 
$O(n' k_j^2 \overline{n} \log \overline{n})=O(n k_j^2 \log \overline{n})$.

\begin{remark}\rm
We consider how many uses of the channel can achieve the capacity when 
$d$ is a prime power $q$
and the intermediate node cannot use additional random number, i.e., $\gamma_i=0$.
To answer this problem, we consider another proof in this special case.
When we set $n':= k_2 \cdots k_c$ and $n:= n'  \cdot  \max_{1 \le i \le c}n_{k_i,r_i} $,
we can achieve the capacity in the following way.
That is, our transmission can be regarded as 
$n'$ times transmission on $ \FF_{q'} $, i.e.,
$n'  \cdot  \max_{1 \le i \le c}n_{k_i,r_i}$ times transmission of the original channel.

In the following construction, 
we employ $ k_1 \cdots  k_c$ random variables.
In this protocol, we securely transmit 
$ (k_1-r_1) \cdots (k_i- r_{i}) k_{i+1} \cdots k_c$ random variable to 
the $i$-th node.
That is,
in the transmission from the $i-1$-th node to the $i$-th node,  
we transmit 
$ (k_1-r_1) \cdots  (k_{i-1}- r_{i-1}) r_{i} k_{i+1} \cdots k_c$ random numbers, in which,
$ (k_1-r_1) \cdots (k_i- r_{i}) k_{i+1} \cdots k_c$  random numbers
are securely transmitted and 
the remaining $ (k_1-r_1) \cdots  (k_{i-1}- r_{i-1}) r_{i} k_{i+1} \cdots k_c$
random variables are treated as scramble random variables.
Such a transmission is possible by applying the method given Subsection \ref{SS1}
to the $ (k_1-r_1) \cdots (k_i- r_{i}) k_{i+1} \cdots k_c$ random variables,
which are securely transmitted to the $i-1$-th node.
Using the above recursive construction, 
we can securely transmit $ \prod_{i=1}^c(k_i-r_i)$ random variables.

The single use of the channel between the $i-1$-th node and the $i$-th node
can securely transmit $ (k_i-r_i)$ random variables.
So, to realize this code, we need to use 
the channel between the $i-1$-th node and the $i$-th node 
at $ \frac{ (k_1-r_1) \cdots (k_i- r_{i}) k_{i+1} \cdots k_c}{k_i-r_i}
= (k_1-r_1) \cdots (k_{i-1}- r_{i-1}) k_{i+1} \cdots k_c$ times. 
That is, to realize this code, we need to use this relay channel
$\max_{1\le i \le c} (k_1-r_1) \cdots (k_{i-1}- r_{i-1}) k_{i+1} \cdots k_c$ times. 
Overall, this code can transmit
\begin{align}
&\min_{1\le i \le c}
\frac{\prod_{i=1}^c (k_i-r_i)}{(k_1-r_1) \cdots (k_{i-1}- r_{i-1}) k_{i+1} \cdots k_c}
\nonumber \\
=&
\min_{1 \le j \le c}
(k_j-r_j) 
\frac{(k_{j+1}-r_{j+1}) \cdots (k_c- r_c) }{k_{j+1} \cdots k_c }
\end{align}
variables per single use of the relay channel.
That is, the transmission rate of this code is
$
\log q \min_{1 \le j \le c}
(k_j-r_j) 
\frac{(k_{j+1}-r_{j+1}) \cdots (k_c- r_c) }{k_{j+1} \cdots k_c }
$.
Therefore, we can realize a code to satisfy the conditions \eqref{Hd1} and \eqref{Hd2} 
for the above given $n$.
\hfill $\square$\end{remark}

\subsection{Scalar linearity}\Label{R2}
Now, we show that this capacity cannot be attained under the scalar linearity condition. 
That is, we consider the special case to satisfy the following conditions.
The intermediate node cannot use additional random number, i.e., $\gamma_i=0$.
We can transmit only a single symbol of a finite filed $\FF_{q'}$ in each channel.
The coding operations are limited to linear operations over the finite filed $\FF_{q'}$.
Since each channel can send only a scalar in $\FF_{q'}$,
this kind of linearity is called the {\it scalar linearity}\cite{Dougherty}.
To distinguish the condition (C1) from the scalar linearity,
the condition (C1) is often called the {\it vector linearity}\cite{Dougherty}.
Existing studies employ one of these constraints as Table \ref{TX}.
Only a deterministic attack is allowed to the eavesdropper.
Under the above condition, 
the number of symbols transmitted securely is not greater than $\max (k_1-\sum_{j=1}^c r_j,0)$, 
which can be shown as follows.

Due to the network structure, the sender can transmit only $k$ symbols $M_1, \ldots, M_k$
in $\FF_{q'}$, where 
the $k$ symbols $M_1, \ldots, M_k$ is given as linear functions of the message and the scramble random variable.
First, we fix the linear coding operation on each nodes. 
In the first group of edges, Eve chooses $r_1$ edges such that 
the information on the $r_1$ edges are given as
$\sum_{i=1}^{r_1} t_{1,i',i }M_i$ with $i'=1,\ldots,r_1$
and $ \{ \vec{t}_{1,i'}\}$ is linearly independent, where $\vec{t}_{1,i'}=(t_{1,i',i})_{i=1}^{k}$ for $i'=1,\ldots,r_1$.
Similarly, 
when $k_1 \ge r_1+r_2 $,
in the second group of edges, Eve chooses $r_2$ edges such that 
the information on the $r_2$ edges are given as
$\sum_{i=1}^{r_2} t_{2,i',i }M_i$ with $i'=1,\ldots,r_1$
and $ \{ \vec{t}_{1,i'}\} \cup \{ \vec{t}_{2,i'}\}$ is linearly independent, where $\vec{t}_{2,i'}=(t_{2,i',i})_{i=1}^{k}$ for $i'=1,\ldots,r_2$.
When $k_1 < r_1+r_2 $,
in the second group of edges, Eve chooses $k_1-r_1$ edges such that 
the information on the $k_1-r_1$ edges are given as
$\sum_{i=1}^{k_1-r_1} t_{2,i',i }M_i$ with $i'=1,\ldots,r_1$
and $ \{ \vec{t}_{1,i'}\} \cup \{ \vec{t}_{2,i'}\}$ is linearly independent, where $\vec{t}_{2,i'}=(t_{2,i',i})_{i=1}^{k}$ for $i'=1,\ldots,k_1-r_1$.
When $k_1 > r_1+r_2 $,
we repeat this process up to the $c$-th group or 
$j'$-th group satisfying $k_1-\sum_{j=1}^{j'} r_j\le 0$.
Hence, the information with dimension $\max(k_1, \sum_{j=1}^c r_j)$ is leaked to the eavesdropper.
Therefore, the number of symbols transmitted securely is not greater than $\max (k_1-\sum_{j=1}^c r_j,0)$.

This fact shows the following effect.
To achieve the capacity even with deterministic attacks,
each channel needs to transmit several symbols in the finite field $\FF_{q'}$.
That is, we need to handle the vector space over the finite field $\FF_{q'}$.
Furthermore, as a special case,
in the setting given in Section \ref{S5},
we find that we need to introduce a non-linear code to realize the situation that 
Eve cannot recover the message perfectly with deterministic attack.

We often increase the size $q'$ of finite field $\FF_{q'}$
in the scalar linearity while
we fix the size $q$ of finite field $\FF_{q}$ and increase the dimension of the vector space in the vector linearity.
In the real communication, the data is given as a sequence of 
$\FF_{2}$.
In this case, when $q=2$, the coding operation satisfying  
the vector linearity can be easily implemented
because the vector linearity reflects the structure of the data.
However, the coding operation satisfying  
the scalar linearity cannot be easily implemented
unless $q'$ is a power of $2$ because 
the scalar linearity does not reflect the structure of the data.
Only when $q'$ is a power of $2$,
the scalar linearity not be easily implemented.
However, even in this case, 
the scalar linearity has worse performance than the vector linearity
due to the above discussion
because the scalar linearity introduces a constraint that does not appear in the vector linearity.
Hence, it is better to impose the vector linearity.

\section{Important lemmas}\Label{S7}
Here, for the latter discussion, we prepare important lemmas.
We denote the set $\{1, \ldots, k\}$ by $[k]$,
and denote the collection of subsets $S\subset [k]$ with cardinality 
$r$ by ${ [k] \choose r}$.

Now, we consider the random variables
$X,\vec{Y}_1, \ldots, \vec{Y}_k$.
For any subset $S \subset [k]$, we denote the tuple of random variables
$(\vec{Y}_{s})_{s \in S}$ by $\vec{Y}_S$.
We can show the following two lemmas.

\begin{lemma}\Label{LH10}
We have
\begin{align}
\sum_{S \in { [k] \choose r}}
H(\vec{Y}_{S}| X)
\ge &
{ k-1 \choose r-1}
H(\vec{Y}_{[k]}| X)
\nonumber \\
=&
\frac{r}{k}{ k \choose k-r}
H(\vec{Y}_{[k]}| X).
\label{KK1}
\end{align}
\hfill $\square$\end{lemma}

\begin{remark}\rm
Lemma \ref{LH10} is known as Han's inequality \cite{Han}, and 
it can be shown by using Baranyai's Theorem \cite{Bara}.
However, this paper shows Lemma \ref{LH10} by using our invented lemma, Lemma \ref{Lemma-1}.
\hfill $\square$\end{remark}

\begin{lemma} \Label{Lemma-1}
Let ${\cal S}_h$ be a collection of subsets of $[k]$.
When any element of $[k]$ is contained in exactly $h$ members of ${\cal S}_h$,
we have
\begin{equation} \Label{eq:Lemma-11}
\sum_{S \in {\cal S}_h} H(\vec{Y}_S|X) \ge h H(\vec{Y}_{[k]}|X).
\end{equation}
\hfill $\square$\end{lemma}

\begin{proofof}{Lemma \ref{Lemma-1}}
We prove the lemma by induction in $h$. When $h=1$, it is trivial. 
Assume that the lemma holds for $h-1$. 
We pick a subcollection ${\cal S}':=\{S_1,S_2, \ldots,S_f\} \subset {\cal S}_h$
such that $\cup_{i=1}^f S_i= [k]$.
We define $S_i' :=S _i \cap (\cup_{j=1}^{i-1}S_j)$ and
${\cal S}_{h-1}=({\cal S}_h\setminus {\cal S}') \cup 
\{S_2', \ldots, S_f'\}$.

We can see that any element of $[k]$ is contained in exactly $h-1$ members of ${\cal S}_{h-1}$, from the following lines. 
Assume that an element $a \in [k]$ is contained in exactly $b$ members of 
${\cal S}'$.
Notice that $a$ is contained by $S_i$, for each particular $i$, if and only if it is contained by exactly one of $S_i\setminus[\cup_{j=1}^{i-1}S_j]$ and $S_i'$. 
For any element $a\in [k]$, there uniquely exists an integer $i$ such that
$a\in S_i\setminus[\cup_{j=1}^{i-1}S_j]$.
So, the element $a$ is contained in exactly $b-1$ members of 
$\{S_2', \ldots, S_f'\}$. 
Therefore, 
the element $a$ is contained in exactly $h-b+(b-1)=h-1$ members of 
${\cal S}_{h-1}=({\cal S}_h\setminus {\cal S}') \cup 
\{S_2', \ldots, S_f'\}$.

Therefore,
\begin{align}
&
\sum_{S \in {\cal S}_h} H(\vec{Y}_S|X) 
=
\sum_{S \in {\cal S}_h\setminus {\cal S}'} H(\vec{Y}_S|X) 
+
\sum_{i=1}^f H(\vec{Y}_{S_i}|X) \nonumber \\
=&
\sum_{S \in {\cal S}_h\setminus {\cal S}'} H(\vec{Y}_S|X) 
+
\sum_{i=1}^f H( \vec{Y}_{S_i'}|X) \nonumber \\
&+
\sum_{i=1}^f H(\vec{Y}_{S_i \setminus S_i'}|   \vec{Y}_{S_i'}X) \nonumber \\
\stackrel{(a)}{\ge} &
\sum_{S \in {\cal S}_{h-1}} H(\vec{Y}_S|X) 
+
\sum_{i=1}^f H(\vec{Y}_{S_i \setminus (\cup_{j=1}^{i-1}S_j)}|   \vec{Y}_{\cup_{j=1}^{i-1}S_j}X) \nonumber \\
 \stackrel{(b)}{\ge}& 
(h-1) H(\vec{Y}_{[k]}|X)+  H(\vec{Y}_{[k]}|X)
=h H(\vec{Y}_{[k]}|X),
\end{align}
where $(a)$ follows from the relation 
$S_i'\subset \cup_{j=1}^{i-1}S_j$
and $(b)$  follows from 
the relation $\cup_{i=1}^f S_i=[k] $
and the induction hypothesis, the fact that
$\sum_{S \in {\cal S}_{h-1}} H(\vec{Y}_S|X) \ge (h-1) H(\vec{Y}_{[k]}|X)$.
\end{proofof}

\begin{proofof}{Lemma \ref{LH10}}
Now, we show Lemma \ref{LH10} by using Lemma \ref{Lemma-1}.
Any element $a \in [k]$ is contained in exactly ${k-1 \choose r-1}  $ members of 
$ {[k] \choose r}$.
So, we apply Lemma \ref{Lemma-1} 
to the case with ${\cal S}_h= {[k] \choose r}$ 
and $h={k-1 \choose r-1} $.
Hence, we have Eq. \eqref{KK1}.
\end{proofof}

\section{homogeneous multicast relay network} \Label{S8}
\subsection{Formulation and capacity regions}\Label{multi1}
Next, 
as a special case of Example \ref{Ex2}, 
we consider the homogeneous multicast relay network (Fig. 4) defined as follows.
This network has one source node and $b$ terminal nodes.
It has $c-1$ groups of intermediate nodes.
The $i$-th group has $b_i$ intermediate nodes, and
the set of $b$ terminal nodes is regarded as the $c$-th group,
and the source node is regarded as the $0$-th group.
So, the numbers $b_0$ and $b_c$ are defined to be $1$ and $b$.
Each node of the $i$-the group is expressed as $n(i,1), \ldots, n(i,b_i)$.

Each node of the $i-1$-th group is connected to every node of the $i$-th group with $k_i$ edges.
That is, there are $b_{i-1}b_i k_i $ edges from the $i-1$-th group to the $i$-th group.
For each node of the $i$-th group,
Eve is assumed to wiretap $r_i$ edges among $b_{i-1} k_i $ edges connected to the node of the $i$-th group
from nodes of the $i-1$-th group.
That is, Eve wiretaps $r_i b_i$ edges among $b_{i-1}b_i k_i $ edges
between the $i-1$-th group and the $i$-th group. 

\begin{figure*}[t!]
\begin{center}
\includegraphics[scale=0.5]{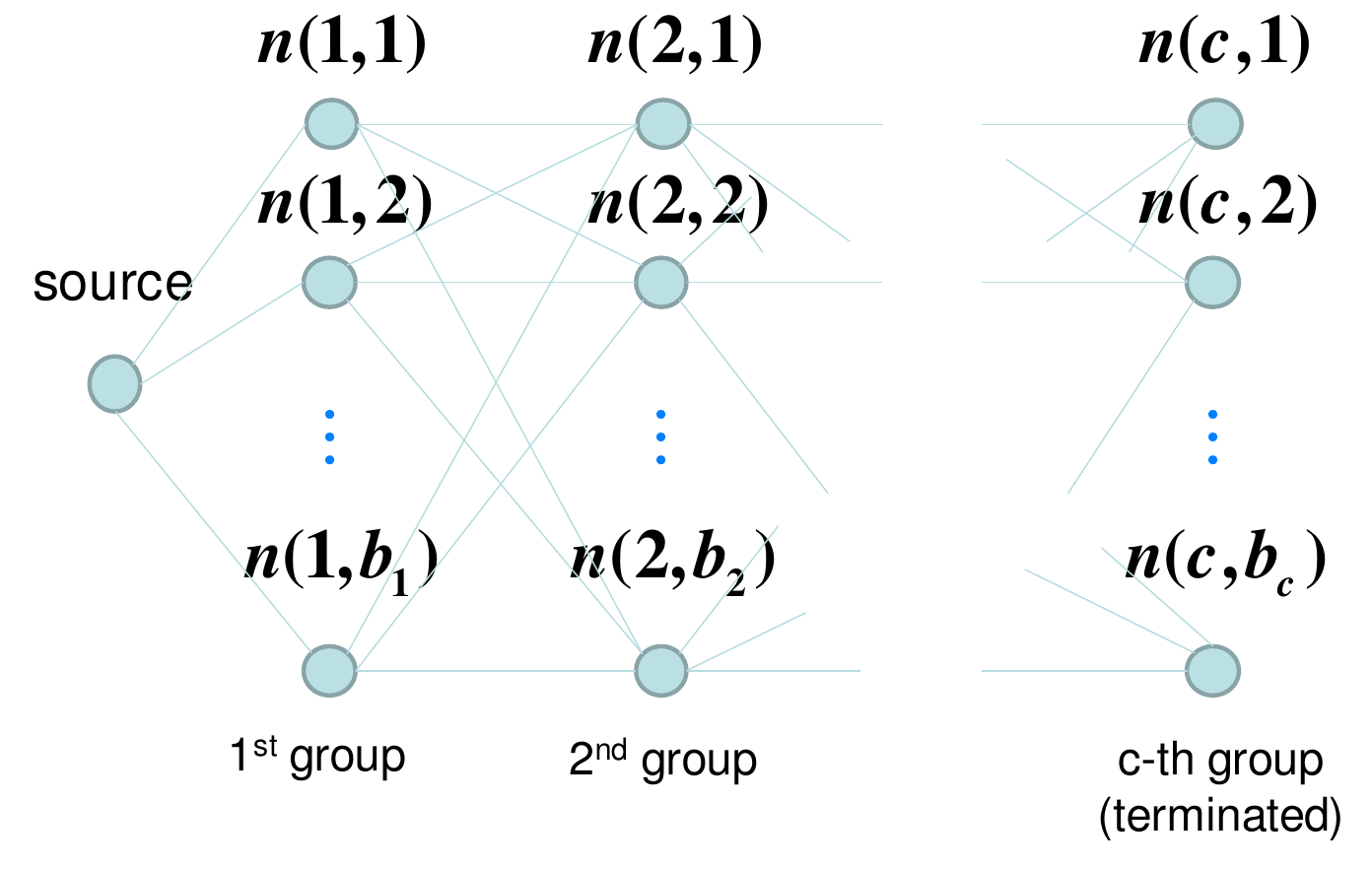}
\end{center}
\caption{homogeneous multicast relay network}
\Label{FT2}
\end{figure*}%

Then, we have the following theorem for the no-randomness capacity region.
\begin{theorem}\Label{TH10}
\begin{align}
&{\cal C}_2= 
{\cal C}_{2,L}= 
{\cal C}_{2,D}= 
{\cal C}_{2,L,D}= 
{\cal C}_{2,AC}= 
{\cal C}_{2,L,AC}\nonumber \\
=& 
\bigg\{(R_1, \ldots, R_b) \bigg| \sum_{i'=1}^b R_{i'} \le A_1,
R_i \le A_2 \hbox{ for } i=1,\ldots, b \bigg\}\Label{e5-6T}
\end{align}
where
\begin{align}
A_1:=&(\log d) \min_{1 \le j \le c}
\bigg((b_{j-1} k_j -r_j) b_j\nonumber \\
&\cdot \frac{(b_j k_{j+1}  -r_{j+1}) \cdots (b_{c-1}k_c - r_c) }{b_j k_{j+1}  \cdots b_{c-1} k_c  } \bigg),\Label{5-10A}\\
A_2:=&
(\log d )(b_{c-1} k_c -r_c) .\Label{5-10B}
\end{align}
\hfill $\square$\end{theorem}

For the full-randomness capacity region, we have the following theorems.
\begin{theorem}\Label{TH8}
Assume that $c=2$ and $r_2/k_2 $ is an integer.
Then, we have
\begin{align}
& {\cal C}_{1}={\cal C}_{1,L}= {\cal C}_2={\cal C}_{2,L}
=
{\cal C}_{1,D}={\cal C}_{1,L,D}= {\cal C}_{2,D}
\nonumber \\
=& {\cal C}_{2,L,D}
=
{\cal C}_{1,AC}={\cal C}_{1,L,AC}= {\cal C}_{2,AC}={\cal C}_{2,L,AC}.
\end{align}
\hfill $\square$\end{theorem}

\begin{theorem}\Label{TH9}
Assume that $c=3$ and $r_3/k_3 $ is an integer.
\begin{align}
&{\cal C}_1={\cal C}_{1,L}
={\cal C}_{1,D}={\cal C}_{1,L,D}
={\cal C}_{1,AC}={\cal C}_{1,L,AC}
\nonumber \\
= &
\bigg\{(R_1, \ldots, R_b) \bigg| \sum_{i'=1}^b R_{i'} \le A_3,
R_i \le A_2 \hbox{ for } i=1,\ldots, b \bigg\}\Label{e5-8},
\end{align}
where
\begin{align}
&A_3\nonumber \\
:=& (\log d) 
\min
\Bigg( (k_1 -r_1) b_1,\nonumber \\
&\min_{2 \le j \le 3} 
(b_{j-1} k_j -r_j) b_j 
\frac{(b_j k_{j+1}  -r_{j+1}) \cdots (b_{c-1}k_c - r_c) }{b_j k_{j+1}  \cdots b_{c-1} k_c  }
\Bigg) \nonumber \\
=&(\log d) 
\min
\Bigg( (k_1 -r_1) b_1,
(b_{1} k_2 -r_2) b_2
\frac{b_{2}k_3 - r_3 }{b_{2} k_3  },
\nonumber \\
&
(b_{2} k_3 -r_3) b_3
\Bigg) .
\end{align}
\hfill $\square$\end{theorem}

\subsection{Converse part for Theorem \ref{TH10}}\Label{S4-2}
We consider the $j$-th group as one intermediate node,
and the set of the $b$ terminal nodes as one terminal node,
which yields a relay network.
Then, applying the relation \eqref{Hd2} to this relay network, 
we obtain the condition $\sum_{i'=1}^b R_{i'} \le A_1 $.

Next, we consider the $j$-th group as one intermediate node,
and focus only on the $i$-th terminal nodes,
which yields another relay network.
Then, applying the relation \eqref{Hd2} to this relay network, 
we obtain the other condition $R_{i} \le A_2 $.

\subsection{Code construction for Theorem \ref{TH10}}\Label{S4-3}
Here, by induction, we make a linear code to achieve the RHS of \eqref{e5-8} when $d$ is a prime power $q$.
In the general case, we can construct the desired linear code by using the method in Lemma \ref{LH3}.
The liner code construction with $c=1$ is given from the code given in Subsection \ref{SS1}.
We construct the desired linear code by induction with respect to the number $c$.

Assume that $n$ is a multiple of $\overline{n}:=\max_{1\le i \le c}n_{b_{i-1}k_i ,r_i}$.
Now, we assume that the source node can securely transmit 
$\sum_{i'=1}^{b} N_{i'}$ letters to each intermediate node in the $c-1$-th group
by $n$ use of the channel.
When $N_{i'} \le n k_c$, under this assumption, 
we can transmit $N_{i'} b_{c-1} -n r_c$ letters from the source node to the $i'$-th terminal node by $n$ use of the channel as follows.
Such a code will be called Code $(N_1 ,\ldots, N_b)$.

For $j_2= 1, \ldots, b$, $j_1= 1, \ldots, k_c$,
we denote 
the $\sum_{i'=1}^{j_2-1}N_{i'} + j_1$-th securely transmitted letter to $j$-th intermediate node in the $c-1$-th group
by $X_{j_2, j_1 + j N_{i'}}$.
Then, for a given $j_2=1, \ldots, b$, the source node
prepares messages 
$M_{j_2, j_3 }$ for $j_3=1, \ldots ,N_{j_2} b_{c-1} -n r_c$
and scramble random numbers
$L_{j_2, j_3 }$ for $j_3=1, \ldots , n r_c$.
Then, the source node
makes conversion from
the pair of $\vec{M}_{j_2} $ and $\vec{L}_{j_2}$ to $\vec{X}_{j_2}$
such that 
there is no information leakage for 
$\vec{M}_{j_2} $ even when any $nr_c$ letters of $\vec{X}_{j_2}$ are eavesdropped.
Such a code can be constructed by using the discussion in 
Subsection \ref{SS1}.

Now, we employ the assumption of induction.
So, there exist an integer $n$ and a code $\Phi_n$ with block-length $n$
such that the rate tuple is 
$(\frac{A_4}{b_{c-1}}, \ldots, \frac{A_4}{b_{c-1}})$, where
\begin{align}
A_4:= & 
\min_{1 \le j \le c-1}
\bigg((b_{j-1} k_j -r_j) b_j \nonumber \\
&\cdot \frac{(b_j k_{j+1}  -r_{j+1}) \cdots (b_{c-2}k_{c-1} - r_{c-1}) }{b_j k_{j+1}  \cdots b_{c-2} k_{c-1}  }\bigg) .
\Label{5-10C}
\end{align}
Using this fact, we show the desired statement by classifying two cases.

(1) Case of $\frac{A_4}{b_{c-1}}\ge  k_c b_c$:
In this case, the minimum in \eqref{5-10A} is realized with $j=c$, which implies $A_1=b_c A_2$.
To attain the RHS of \eqref{e5-8},
it is sufficient to give a code with the rate tuple 
$( A_2, \ldots, A_2)=
( \log d (b_{c-1} k_c-r_c), \ldots, \log d (b_{c-1} k_c-r_c))$.
The required secure transmission from the source node to the $c-1$-th group
is possible as follows.
Combining the assumption of induction and Code 
$(n k_c ,\ldots, n k_c)$.
We obtain a linear code with the rate tuple 
$
( \log d (b_{c-1} k_c-r_c), \ldots, \log d (b_{c-1} k_c-r_c))$.

(2) Case of $\frac{A_4}{b_{c-1}}<  k_c b_c$:
We have $A_1= A_4 \frac{ b_{c-1}k_{c} - r_{c}}{b_{c-1} k_c}$.
To attain the RHS of \eqref{e5-8},
it is sufficient to give a code with the rate tuple 
$( R_1, \ldots, R_b)$ satisfying conditions
$\sum_{i'=1}^b R_{i'}\le A_1$ and $R_i \le A_2$.
Due to the assumption of induction,
the source node can securely transmit 
$n \frac{A_4}{b_{c-1}} $ letters to each node in the $c-1$-th group.
Now, we choose $n$ such that $n \frac{A_4}{b_{c-1}b }$ is an integer,
$n$ is a multiple of $\overline{n}$,
and $n R_{i'}$ is integer for $i'=1, \ldots, b$.
Therefore, using Code $(n R_{1}, \ldots, n R_{b})$,
we obtain a linear code, in which, 
 the source node can securely transmit to the $i'$-th terminal 
with rate $R_{i'}$.
Since this linear code construction requires only the conditions
$\sum_{i'=1}^b R_{i'}\le A_1$ and $R_i \le A_2$,
the RHS of \eqref{e5-8} is attained.

Due to the same discussion as Section \ref{SS1}
node operations on node of $i$-th group has complexity 
$O(n k_i^2 \log \overline{n})$.

\subsection{Proof of Theorem \ref{TH8}}\Label{SS7}
To show Theorem \ref{TH8}, it is sufficient to show the converse part, i.e.,
$ {\cal C}_{1,D}\subset {\cal C}_{2,D}$.
The $i$-th intermediate node can transmit information
of $k_2$ symbols per single use of channel to the $j$-th terminal node.
In order that the $j$-th terminal node recovers the original message $M_j$,
the $j$-th terminal node needs to recover 
a part of information $M_{i,j}$ with respect to the original message that
is determined by the information received by the $i$-th intermediate node.
That is, collecting the variables $M_{1,j}, \ldots, M_{b_1,j} $,
the $j$-th terminal node recovers $M_j$.
We choose an injective function $s$ from $\{1, \ldots, r_2/k_2\} $ to
$\{1, \ldots, b_2\} $.
Now, we consider the case that Eve wiretaps all the channels 
from the $s(i)$-th intermediate node to the $j$-th terminal node
for $i=1,\ldots, r_2/k_2$.
When the $s(i)$-th terminal node introduces scramble random variables $L_{s(i),j}$ in the channel to the $j$-th terminal node,
the $j$-th terminal node needs to recover $M_{s(i),j}$.
In this case, Eve also recovers $M_{s(i),j}$.
Then, there is no merit to introduce the scramble random variables $L_{s(i),j}$ in this channel.
When the $i'$-th terminal node introduces scramble random variables $L_{i',j}$ in the channel to the $j$-th terminal node
for $i' \in \{1, \ldots, b_2\} \setminus \{s(1), \ldots, s(r_2/k_2)\}$,
the $j$-th terminal node needs to recover $M_{i',j}$.
In this case, Eve has no access to this channel.
Hence, there is no need to introduce the scramble random variables $L_{i',j}$ in this channel.
Therefore, considering this special case, there is no advantage to introduce 
scramble random variables in the intermediate nodes.
That is, any code can be reduced to a code with the no-randomness condition (C2).

\subsection{Proof of Theorem \ref{TH9}}
Due to the discussion in Subsection \ref{SS7},
the scramble random number introduced in intermediate nodes in the 2nd group does not work.
Hence, we obtain the converse part, i.e., 
${\cal C}_D
\subset
\bigg\{(R_1, \ldots, R_b) \bigg| \sum_{i'=1}^b R_{i'} \le A_3,
R_i \le A_2 \hbox{ for } i=1,\ldots, b \bigg\}$.

Next, we construct a code to achieve the capacity region.
Each intermediate node in the first group can securely transmit 
to each terminal node
with the following capacity region:
\begin{align}
\bigg\{(R_1, \ldots, R_b) \bigg| \sum_{i'=1}^b R_{i'} \le \frac{A_5}{b_1},
R_i \le \frac{A_2}{b_1}
 \hbox{ for } i=1,\ldots, b \bigg\}\Label{e5-6}
\end{align}
with
\begin{align}
& A_5\nonumber \\
:=& (\log d) 
\min
\Bigg( 
(b_{1} k_2 -r_2) b_2
\frac{b_{2}k_3 - r_3 }{b_{2} k_3  },
 (b_{2} k_3 -r_3) b_3
\Bigg) .
\end{align}

Now, the source node can securely transmit information to each intermediate node in the first group with the rate
$(\log d) (k_1 -r_1) $.
Combining these discussions, 
the source node can securely transmit information to each terminal node
via a specific intermediate node in the first group
with the following capacity region:
\begin{align}
\bigg\{(R_1, \ldots, R_b) \bigg| \sum_{i'=1}^b R_{i'} \le \frac{A_3}{b_1},
R_i \le \frac{A_2}{b_1}
 \hbox{ for } i=1,\ldots, b \bigg\}
\end{align}
because $A_3= \min ( (\log d) (k_1 -r_1) b_1,A_5) $.
Summing up the above region with respect to intermediate nodes in the first group,
we find 
the relation ${\cal C}_{D,L}
\supset
\bigg\{(R_1, \ldots, R_b) \bigg| \sum_{i'=1}^b R_{i'} \le A_3,
R_i \le A_2 \hbox{ for } i\!=\!1,\ldots, b \bigg\}$, which is the direct part.

\section{homogeneous multiple multicast relay network} \Label{S9}
\subsection{Formulation and capacity regions}\Label{SS2-1}
Next, as a special case of Example \ref{Ex2}, 
we consider the homogeneous multiple multicast relay network (Fig. 5) defined as follows.
This network has $a$ source nodes and $b$ terminal nodes.
It has $c-1$ groups of intermediate nodes.
The $i$-th group has $b_i$ intermediate nodes, and
the set of $b$ terminal nodes is regarded as the $c$-th group,
and the source node is regarded as the $0$-th group.
So, the numbers $b_0$ and $b_c$ are defined to be $a$ and $b$.
Each node of the $i$-the group is expressed as $n(i,1), \ldots, n(i,b_i)$.

Each source code is connected to each intermediate node in the first group with $k_1$ edges.
For $i \ge 2$,
each node of the $i-1$-th group is connected to every node of the $i$-th group with $k_i$ edges.
That is, there are $b_{i-1}b_i k_i $ edges from the $i-1$-th group to the $i$-th group.
For each node of the $i$-th group,
Eve is assumed to wiretap $r_1$ edges among $k_1 $ edges 
between each source node and each intermediate node in the first group.
Totally, Eve wiretaps $ab_1 r_1$ edges among $ab_1 k_1$ edges between the $0$-th group and the first group.
For $i \ge 2$, Eve is assumed to wiretap $r_i$ edges among $b_{i-1} k_i $ edges connected to the node of the $i$-th group from nodes of the $i-1$-th group.
That is, Eve wiretaps $r_i b_i$ edges among $b_{i-1}b_i k_i $ edges
between the $i-1$-th group and the $i$-th group.

\begin{figure*}[t!]
\begin{center}
\includegraphics[scale=0.5]{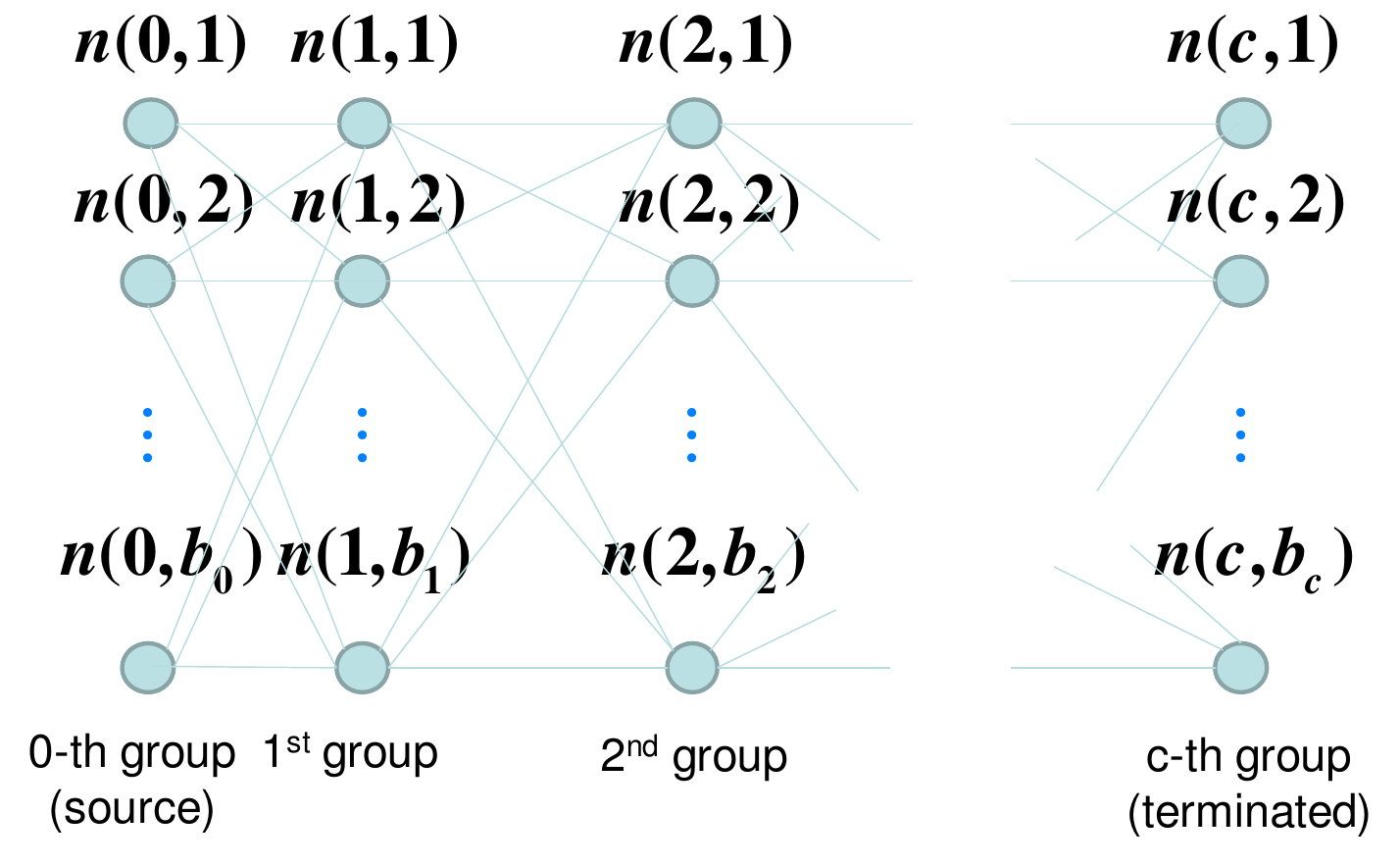}
\end{center}
\caption{homogeneous multiple multicast relay network}
\Label{FT3}
\end{figure*}%

Then, we have the following theorem for the no-randomness capacity region.
\begin{theorem}\Label{TH11}
\begin{align}
&{\cal C}_2= 
{\cal C}_{2,L}= 
{\cal C}_{2,D}= 
{\cal C}_{2,L,D}= 
{\cal C}_{2,AC}= 
{\cal C}_{2,L,AC} 
\nonumber \\
=\! &
\left\{(R_{i,j})_{1\le i \le a, 1\le j \le b} \left| 
\begin{array}{l}
\sum_{i',j'} R_{i',j'} \le B_1,\\
\sum_{j'} R_{i,j'} \le B_2,\\
\sum_{i'}R_{i',j} \le B_3 \\
\hbox{ for } i\!=\!1,\ldots, a , ~j\!=\!1,\ldots, b 
\end{array}
\right.\right\}, \Label{e5-7T}
\end{align}
where
\begin{align}
&B_1
\nonumber \\
:=&(\log d) 
\min\Bigg(
 a (k_1 -r_1) b_1
\frac{(b_1 k_{2} \!-\!r_{2}) \cdots (b_{c-1}k_c \!-\! r_c) }{b_1 k_{2}  \cdots b_{c-1} k_c  } ,
\nonumber \\
&
\min_{2 \le j \le c}
(b_{j-1} k_j -r_j) b_j
\frac{(b_j k_{j+1}  \!-\!r_{j+1}) \cdots (b_{c-1}k_c \!-\! r_c) }{b_j k_{j+1}  \cdots b_{c-1} k_c  } 
\Bigg),
\Label{5-10X}\\
&B_2\nonumber \\
:=&(\log d) 
\min\Bigg(
 (k_1 -r_1) b_1
\frac{(b_1 k_{2}  -r_{2}) \cdots (b_{c-1}k_c - r_c) }{b_1 k_{2}  \cdots b_{c-1} k_c  } ,
\nonumber \\
&
\min_{2 \le j \le c}
(b_{j-1} k_j \!-\!r_j) b_j
\frac{(b_j k_{j+1} \! -\!r_{j+1}) \cdots (b_{c-1}k_c \!-\! r_c) }{b_j k_{j+1}  \cdots b_{c-1} k_c  } 
\Bigg)
\Label{5-11X},
\end{align}
and
\begin{align}
B_3
:=
(\log d )(b_{c-1} k_c -r_c) .\Label{5-10T}
\end{align}
\hfill $\square$\end{theorem}

For the full-randomness capacity region, we have the following theorems.
\begin{theorem}\Label{TH18}
Assume that $c=2$ and $r_2/k_2 $ is an integer.
Then, we have
\begin{align}
&{\cal C}_1= {\cal C}_{1,L}=
{\cal C}_2= {\cal C}_{2,L}
={\cal C}_{1,D}= {\cal C}_{1,L,D}=
{\cal C}_{2,D}\nonumber \\
=& {\cal C}_{2,L,D}
={\cal C}_{1,AC}= {\cal C}_{1,L,AC}=
{\cal C}_{2,AC}= {\cal C}_{2,L,AC}.
\end{align}
\hfill $\square$\end{theorem}

\begin{theorem}\Label{TH19}
Assume that $c=3$ and $r_3/k_3 $ is an integer.
\begin{align}
&{\cal C}_1= {\cal C}_{1,L}= 
{\cal C}_{1,D}= {\cal C}_{1,L,D}
=
{\cal C}_{1,AC}= {\cal C}_{1,L,AC}
\nonumber \\
=& 
\left\{(R_{i,j})_{1\le i \le a, 1\le j \le b} \left| 
\begin{array}{l}
\sum_{i',j'} R_{i',j'} \le B_4,\\
\sum_{j'} R_{i,j'} \le B_5,\\
\sum_{i'} R_{i',j} \le B_3 \\
\hbox{ for } i\!=\!1,\ldots, a , ~j\!=\!1,\ldots, b 
\end{array}
\right. \right\}
\Label{e5-8X},
\end{align}
where
\begin{align}
& B_4 \nonumber \\
:=& (\log d) 
\min
\Bigg( a( k_1 -r_1) b_1,
\nonumber \\
& \min_{2 \le j \le 3}
(b_{j-1} k_j -r_j) b_j
\frac{(b_j k_{j+1}  -r_{j+1}) \cdots (b_{c-1}k_c - r_c) }{b_j k_{j+1}  \cdots b_{c-1} k_c  }
\Bigg) \nonumber \\
=& (\log d) 
\min
\Bigg( a( k_1 -r_1) b_1,
\nonumber \\
&(b_{1} k_2 -r_2) b_2
\frac{b_{2}k_3 - r_3 }{b_{2} k_3  },
(b_{2} k_3 -r_3) b_3
\Bigg), \\
&B_5 \nonumber \\
:=& (\log d) 
\min
\Bigg( (k_1 -r_1) b_1,
\nonumber \\
& \min_{2 \le j \le 3}
(b_{j-1} k_j -r_j) b_j
\frac{(b_j k_{j+1}  -r_{j+1}) \cdots (b_{c-1}k_c - r_c) }{b_j k_{j+1}  \cdots b_{c-1} k_c  }
\Bigg) \nonumber \\
=& (\log d) 
\min
\Bigg( (k_1 -r_1) b_1,\nonumber \\
& (b_{1} k_2 -r_2) b_2
\frac{b_{2}k_3 - r_3 }{b_{2} k_3  },
(b_{2} k_3 -r_3) b_3
\Bigg) .
\end{align}
\hfill $\square$\end{theorem}

\subsection{Converse part for Theorem \ref{TH11}}\Label{SS2-2}
We consider the $j$-th group as one intermediate node,
and the set of the $b$ terminal nodes and
the set of the $a$ source nodes as one terminal node and 
one source node, respectively,
which yields a relay network.
Then, applying the relation \eqref{Hd2} to this relay network, 
we obtain the condition $\sum_{i',j'}^b R_{i',j'} \le B_1 $.

Applying the discussion in Subsection \ref{S4-2} to the network
from the $i$-th source node to the $j$-th group,
we obtain the condition $\sum_{j'}^b R_{i,j'} \le B_2 $.
Similarly, applying the discussion in Subsection \ref{S4-2} to the network
from the first group to the $j$-th terminal node,
we obtain the condition $ \sum_{i'}R_{i',j} \le B_3 $.

\subsection{Code construction for Theorem \ref{TH11}}\Label{SS2-3}
Here, by induction, we make a code to achieve the RHS of \eqref{e5-7T} when $d$ is a prime power $q$.
In the general case, we can construct the desired code by using the method in Lemma \ref{LH3}.
The code construction with $c=1$ is given from the code given in Subsection \ref{SS1}.
We construct the desired code by induction with respect to the number $c$.

Assume that $n$ is a multiple of $\overline{n}:=\max_{1\le i \le c}n_{b_{i-1}k_i ,r_i}$.
We choose a rate tuple $( R_{i,j})_{i,j}$ satisfying the condition in the RHS of \eqref{e5-7T}.
As mentioned in the proof of Theorem \ref{TH10},
when we can securely transmit an unlimited number of messages from the source node to
all of intermediate nodes in the $c-1$-th group,
using the code with block-length $n$ constructed in Subsection \ref{SS1},
we can transmit $n(b_{c-1} k_c- r_c)$ letters 
from the source node to each terminal node, in which,
the source node securely transmits $n k_c$ letters to each intermediate node in the $c-1$-th group.
Therefore, the rate tuple $( R_{i,j})_{i,j}$ can be realized by secure transmission with the rate 
$R_{i,j}':=\frac{  b_{c-1} k_c }{b_{c-1} k_c -r_c } \sum_{j} R_{i,j}$ 
from the $i$-th source node to 
the $j$-th intermediate node in the $c-1$-th group.
The assumption of induction guarantees that the rate tuple $( R_{i,j}')_{i,j}$ is attainable in the network from the first group to the $c-1$-th group
because the rate tuple $( R_{i,j}')_{i,j}$ satisfies the conditions
$\sum_{i',j'} R_{i',j'}' \le B_1',
\sum_{j'} R_{i,j'}' \le B_2',
\sum_{i'}R_{i',j}' \le B_3' $ for $ i=1,\ldots, a , ~j=1,\ldots, b_{c-1} $,
where
\begin{align}
B_1':=&(\log d) \min_{1 \le j \le c-1}
\bigg(
(b_{j-1} k_j -r_j) b_j\nonumber \\
&\cdot \frac{(b_j k_{j+1}  -r_{j+1}) \cdots (b_{c-2}k_{c-1} - r_{c-1}) }{b_j k_{j+1}  \cdots b_{c-2} k_{c-1}  } 
\bigg),\\
B_2':=&(\log d) \nonumber \\
&\cdot \min\Bigg(
 (k_1 -r_1) b_1
\frac{(b_1 k_{2}  -r_{2}) \cdots (b_{c-2}k_{c-1} - r_{c-1}) }{b_1 k_{2}  \cdots b_{c-2} k_{c-1}  } ,
\nonumber \\
&\min_{2 \le j \le c-1}
\bigg(
(b_{j-1} k_j -r_j) b_j \nonumber \\
&\cdot \frac{(b_j k_{j+1}  -r_{j+1}) \cdots (b_{c-2}k_{c-1} - r_{c-1}) }{b_j k_{j+1}  \cdots b_{c-2} k_{c-1}  } 
\bigg)
\Bigg),
\\
B_3':= &
(\log d )(b_{c-2} k_{c-1} -r_{c-1}) .
\end{align}
Therefore, the rate tuple $( R_{i,j})_{i,j}$ is achievable.
Due to the same discussion as Section \ref{SS1}
node operations on node of $i$-th group has complexity 
$O(n k_i^2 \log \overline{n})$.

\subsection{Proof of Theorem \ref{TH18}}\Label{SS17}
To show Theorem \ref{TH8}, it is sufficient to show the converse part
$ {\cal C}_1\subset {\cal C}_2$.
As shown in the proof of Theorem \ref{TH8},
any code can be reduced to a code with the no-randomness condition (C2).
Hence, we obtain $ {\cal C}_1\subset {\cal C}_2$.

\subsection{Proof of Theorem \ref{TH19}}
Similar to the proof of Theorem \ref{TH9},
the scramble random number introduced in intermediate nodes in the 2nd group do not work.
Hence, we obtain the converse part.

Next, we construct a code to achieve the capacity region.
Each source node can securely transmit information to each intermediate node in the first group with the rate
$(\log d) (k_1 -r_1) $.
Combining this code and the codes given in \eqref{e5-6} from each intermediate node in the first group to 
 each terminal node,
the set of source nodes can securely transmit information to each terminal node
via a specific intermediate node in the first group
with the following capacity region:
\begin{align}
\left\{(R_{i,j})_{1\le i \le a, 1\le j \le b} \left| 
\begin{array}{l}
\sum_{i',j'} R_{i',j'} \le \frac{B_4}{b_1},\\
\sum_{j'} R_{i,j'} \le \frac{B_5}{b_1}, \\
\sum_{i'} R_{i',j} \le \frac{B_3}{b_1} \\
\hbox{ for } i=1,\ldots, a , ~j=1,\ldots, b 
\end{array}
\right.\right\}
\end{align}
because $B_3=A_3$,
$B_4= \min ( (\log d) a (k_1 -r_1) b_1,A_5) $
and $B_5= \min ( (\log d) (k_1 -r_1) b_1,A_5) $.
Summing up the above region with respect to intermediate nodes in the first group,
we find that the rate region defined in the RHS of \eqref{e5-8}.

\section{Conclusion}
We have studied active and adaptive attacks,
and have investigated whether an adaptive attack improves Eve's ability.
As our result, we have shown that 
an adaptive attack improves Eve's ability when our code is a linear code.
However, when our code is not a linear code,
we have found an example where an adaptive attack improves Eve's ability
in Section \ref{S5}.
Any linear code cannot realize the performance of the non-linear code given there under the setting of Section \ref{S5}
when Eve is allowed to a deterministic attack.
Hence, the improvement by the adaptive attack is essential in this setting.

Next, we consider several types of network, in which there is restriction for randomness in the intermediate nodes.
This kind of restriction is crucial in the secure network because
randomness is required to realize the secrecy.
In the latter part of this paper, we have addressed various types of relay networks 
in the asymptotic setting, where we employ liner codes, i.e., these codes are given as vector spaces over a finite field.
In Section \ref{S6}, we have considered a typical type of unicast relay network 
and have derived the capacity under various restrictions for randomness in the intermediate nodes.
To show the converse part, we have shown a notable lemma in Section \ref{S7}.
Our proof of the direct part follows from a lemma related to wiretap channel II.
Also, in Subsection \ref{R2}, 
we have shown that the code does not work when it is given as a scalar of a finite field.
Further, we have proceeded to more complicated networks, e.g., 
 a typical type of multicast relay network and  a typical type of multiple multicast relay network.
 Since their asymptotic performances are characterized as their capacity regions,
in Sections \ref{S8} and \ref{S9},
we have derived them under the condition that the intermediate nodes have no scramble random number by generalizing the method used in Section \ref{S6}.

While our asymptotic results are limited to special networks,
the minimum cut theorem does not work in these networks.
Hence, our codes suggest a general theory for networks whose capacity cannot be shown by the minimum cut theorem.
It is an interesting future study to establish such a theory.
As explained in Section \ref{S1},
when the spaces of the intermediate nodes and/or the budget are limited,
it might be better to avoid to equip scramble random variables in the intermediate nodes. 
The study with this constraint is much desired for the practical viewpoint.

\section*{Acknowledgments}
The authors are very grateful to Dr. Wangmei Guo
for her helpful discussions and her hospitality during the authors' stay in Xidian University.
They also are grateful to Mr. Seunghoan Song for helpful comments for Appendix and Lemma 4.
The authors thank a reviewer of the previous version of this paper for 
explaining the network code given in Fig. \ref{FT}.
They are also grateful to the referees of this paper to their helpful comments.

\appendices
\section{Proof of Lemma \ref{L425}}\Label{A1}
When $k=r$, it is trivial. When $k=r+1$, we do not need to make any algebraic extension
because it is sufficient to choose $r$ vectors $v_1, \ldots, v_r \in \FF_{q}^{k}$ 
such that $v_{i,j}$ with $j=1, \ldots, r$ is $\delta_{i,j}$
and $v_{i,r+1}$ is $1$.

Now, we consider the case when $k> r+1 $.
When $q>p$, we choose element $e_1, \ldots, e_t$ such that
$\FF_{q}$ is given as $\FF_p[e_1, \ldots, e_t]$.
When $t < k-2$, we make further algebraic extension 
$\FF_p[e_1, \ldots, e_{k-2}]$ by adding elements $e_{t+1}, \ldots, e_{k-2}$.
Now, we denote $1$ by $e_0$.
Then, we choose $r$ vectors $v_1, \ldots, v_r \in \FF_p[e_1, \ldots, e_{k-2}]^k$ by
\begin{align}
v_{i,j}:=
\left\{
\begin{array}{ll}
\delta_{i,j} & \hbox{ when } j \le r \\
1 & \hbox{ when } j = r+1 \\
e_{i+j-r-2} & \hbox{ when } j > r+1.
\end{array}
\right.
\end{align}
 
We can show that the $r$ vectors $v_1, \ldots, v_r$ 
satisfy the required condition as follows.
Choose the function $s$ such that $s(1)< \ldots< s(r)$.
It is sufficient to show that the vector 
$(v_{1,s(r)}, \ldots, v_{r,s(r)})$
cannot be written as a linear combination of 
$(v_{1,s(1)}, \ldots, v_{r,s(1)}), \ldots,
(v_{1,s(r-1)}, \ldots, v_{r,s(r-1)})$.
When $s(r)=r$ or $r+1$, it is trivial.
So, we show the case when $s(r)>r+1$.
Since all entries of $v_{i,s(j)}$ belong to $\FF_p[e_1, \ldots, e_{s(r)-2}]$,
we choose coefficients $\alpha_1, \ldots, \alpha_r\in  \FF_p[e_1, \ldots, e_{s(r)-2}]$
such that $\sum_{i=1}^r \alpha_i v_{j,s(i)}=0$ for $j=1,\ldots, r$.
We show the desired statement by assuming $\alpha_r=1$.

For $i=1, \ldots, r-1$, we divide the coefficient $\alpha_i$ into $r$ parts, i.e., 
we choose $\alpha_{i,j}\in \FF_p[e_1, \ldots, e_{s(r)+j-r-2}]\setminus 
\FF_p[e_1, \ldots, e_{s(r)+j-r-3}]$ as 
$\alpha_i=\sum_{j=1}^r \alpha_{i,j}$.
Since we have $\sum_{i=1}^{r-1} \alpha_i v_{j,s(i) } = -  e_{s(r)+j-r-2} 
\in \FF_p[e_1, \ldots, e_{s(r)+j-r-2}]$ for $j=1,\ldots, r$,
we have
$\sum_{i=1}^{r-1} \alpha_{i,j'} v_{j,s(i) } =0$
for $j' >j $ because $\alpha_{i,j'} \notin \FF_p[e_1, \ldots, e_{s(r)+j-r-2}]$
and $v_{j,s(i) } \in \FF_p[e_1, \ldots, e_{s(r)+j-r-2}]$.

That is, the vectors 
$\bm{\alpha}(j'):= (\alpha_{1,j'}, \ldots, \alpha_{r-1,j'})^T $ for $j'=1, \ldots, r$
and
$\bm{\beta}(j):= (v_{j, s(1) }, \ldots, v_{j,s(r-1) })^T $ for $j=1, \ldots, r-1$
satisfy the conditions:
\begin{align}
(\bm{\alpha}(j'),\bm{\beta}(j)) &=0 \hbox{ for } j'>j \\
\sum_{j'=1}^j (\bm{\alpha}(j'),\bm{\beta}(j)) &=-e_{s(r)+j-r-2}.
\end{align}
Since $\sum_{j'=1}^{j-1} (\bm{\alpha}(j'),\bm{\beta}(j)) \in 
\FF_p[e_1, \ldots, e_{s(r)+j-r-3}]$,
and $-e_{s(r)+j-r-2} (\neq 0) \notin \FF_p[e_1, \ldots, e_{s(r)+j-r-3}]$,
we have $(\bm{\alpha}(j),\bm{\beta}(j))\neq 0$  for $j=1,\ldots, r-1$. 
These properties are summarized as 
\begin{align}
(\bm{\alpha}(i),\bm{\beta}(j))
\left\{
\begin{array}{ll}
= 0 & \hbox{ when } i > j \\
=\neq 0
 & \hbox{ when } i =j .
 \end{array}
 \right.
 \end{align}
The property of triangle matrix implies that 
$\bm{\beta}(1), \ldots, \bm{\beta}(r-1) $ 
are linearly independent.
Since $(\bm{\alpha}(r),\bm{\beta}(j))=0$ for $j=1, \ldots, r-1$,
and $\alpha(r)$ is a $r-1$-dimensional vector,
we have $\bm{\alpha}(r)=0$, which implies 
 $ e_{s(r)-2}=0$. So, we obtain contradiction.


\end{document}